\documentclass{jfm}

\usepackage{url}
\usepackage{natbib}
\usepackage{amsmath}
\usepackage{color}
\usepackage{psfrag}
\usepackage[dvipsnames]{xcolor}
\usepackage{ulem}
\usepackage{comment}
\usepackage{graphicx}
\usepackage[utf8]{inputenc}
\usepackage{mathabx}
\usepackage{multirow}
\usepackage{datetime}
\def\figdirp{./}
\newcommand{\opencircs}{\mbox{$\circ$}}
\newcommand{\opentriangles}{\mbox{\scriptsize $\vartriangle$}}
\newcommand{\blue}[1]{{\color{blue}$1}}
\newcommand{\pline}[3]{\ensuremath{#1\hspace{-0.3cm}\protect\rule[#2ex]{.4cm}{#3pt}}} 
\newcommand{\bline}[2]{\protect\mbox{\rule[#1ex]{.13cm}{#2pt}}}
\newcommand{\lineline}[2]{\protect\mbox{\rule[0.5ex]{#1cm}{#2pt}}}

\def\ocircline{\pline{\raisebox{0.18ex}{\,\,\opencircs}}{0.75}{.4}}
\newcommand{\sfullcirc}{\mbox{{$\bullet$}}}
\newcommand{\tinyfullcirc}{\mbox{{\scriptsize $\bullet$}}}
\def\ccircline{\,\pline{\raisebox{0.28ex}{\,\tinyfullcirc}}{0.75}{.4}}
\def\ccirclinet{\pline{\raisebox{0.14ex}{\,\,\sfullcirc}}{0.75}{.7}}
\def\triline{\,\pline{\raisebox{0.3ex}{\opentriangles}}{0.75}{0.4}}

\def\linedash{\ensuremath{\bline{0.5}{.4}\,\bline{0.5}{.4}}}
\def\linedasht{\ensuremath{\bline{0.5}{.7}\,\bline{0.5}{.7}}}
\def\thinline{\lineline{.3}{.4}}
\def\thickline{\lineline{.3}{.7}}
\def\bthinline{{\color{blue}\thinline}}
\def\bthickline{{\color{blue}\thickline}}
\def\blinedash{{\color{blue}\linedash}}
\def\blinedasht{{\color{blue}\linedasht}}

\def\geomfac{\ensuremath{\frac{ak}{2}\exp(-k\xi)}}
\def\utau{\ensuremath{u_\tau}}
\def\utauv{\ensuremath{\hat{u}_\tau}}
\newcommand\mode[2]{\hat{#1}_{#2}}
\def\p{\partial}
\def\ie{{\em{i.e.\,}}}
\def\dy{\ensuremath{\mathcal{D}}}

\def\L{\ensuremath{\mathcal{L}}}
\def\ut{\ensuremath{u_{\tau}}}

\def\us{u_{i}^{\psi}}
\def\uh{u_{i}^{h}}
\def\up{u_{i}^{p}}
\def\upj{u_{j}^{p}}
\def\uhj{u_{j}^{h}}
\def\upot{u_{i}^{\phi}}
\def\usj{u_{j}^{\psi}}
\def\upotj{u_{j}^{\phi}}

\def\vho{\hat{v}_1^{h,\mathrm{im}}}
\def\vpo{\hat{v}_1^{p,\mathrm{im}}}
\def\vo{\hat{v}_1^{im}}
\def\ppo{\hat{p}_1^{p,\mathrm{re}}}

\def\wuu{\widehat{uu}}
\def\wuv{\widehat{uv}}
\def\tuu{\widehat{\av{u'u'}}}
\def\tuv{\widehat{\av{u'v'}}}

\def\lp{\ensuremath{\nu/\ut}}
\def\lpi{\ensuremath{\ut/\nu}}

\def\cp{\ensuremath{c/\ut}}
\def\retau{\ensuremath{Re_\tau}}
\def\rew{\ensuremath{Re_w}}

\newcommand{\av}[1]{\langle#1\rangle}

\newcommand{\mderiv}[2]{\frac{\p #1}{\p #2}}
\newcommand{\mmderiv}[2]{\frac{\p^2 #1}{\p #2^2}}
\newcommand{\mnderiv}[3]{\frac{\p^2 #1}{\p #2\p #3}}
\newcommand{\deriv}[2]{\p_{#2}#1}
\newcommand{\ccolvec}[3]{  \begin{bmatrix}  #1 \\ #2 \\ #3 \end{bmatrix}  }
\newcommand*\dint{\mathop{}\!\mathrm{d}}

\title{The role of wave kinematics in turbulent flow over waves}
\author{Espen Åkervik and Magnus Vartdal} \affiliation{ Norwegian
  Defence Research Establishment (FFI), P.O. Box 25, NO-2027 Kjeller,
  Norway}

\begin{document}
\maketitle
\begin{abstract}
The turbulent flow over monochromatic waves of moderate steepness is
studied by means of wall resolved large eddy simulations.  The
simulations cover a range of wave ages for several Reynolds numbers.
We compute the Fourier modes of the flow variables and analyse the
momentum balance for the mean and fundamental mode, with the primary
goal of understanding the dependence of form drag on the governing
parameters.  At low wave ages, the form drag displays a large
sensitivity to changes in Reynolds number, and the interaction between
turbulent and wave-induced stresses increases with Reynolds number.
At higher wave ages, the flow enters a quasi-laminar regime where the
wave-induced stress is primarily balanced by viscous stresses.
  
To exploit the increasing importance of the wave kinematics observed
in the intermediate to high wave age regime, a novel split system
approach is introduced.  The key ingredient in this approach is that
in the absence of background shear, the flow response to the wave
kinematics is laminar. This laminar solution acts as a forcing in a
RANS type model for the shear flow that is subject to homogeneous
boundary conditions.  To account for the effects of turbulence, we
force the system using Reynolds stresses from the corresponding large
eddy simulation.  We give an analytic functional dependence for the
form drag associated with the laminar solution.  For intermediate to
high wave ages, the form drag of the shear flow exhibits relatively
simple behaviour, and we derive approximate functional dependencies
for the quasi-laminar regime.
  
The high sensitivity of the form drag to variation in Reynolds number
at low wave ages is more challenging to evaluate using the split
system approach. This is primarily due to the increased importance of
nonlinearity in the shear flow. These nonlinearites are inherently
coupled with higher harmonics in the turbulent stresses. Nevertheless,
the split system approach can be utilized to quantify the importance
of different harmonics in the turbulent stresses by explicitly
choosing which modes to include in the split system forcing. We
demonstrate that the fundamental mode of the Reynolds stresses becomes
increasingly important for accurate prediction of the form drag as the
Reynolds number increases.

\end{abstract}

\section{Introduction}
Stresses from turbulent wind over water generate waves, which in turn
alter the wind. The exchange of momentum, energy and mass occurring at
air-water interfaces is to a large extent determined by this wind-wave
interaction. Despite receiving constant attention from the research
community in the form of field measurements
\citep{hasselmann1973,hristov1998,edson2007}, laboratory experiments
\citep{plant1982,mastenbroek1996,grare2013,buckley2016}, theoretical
studies \citep{phillips1957,miles1957,janssen1991,belcher1993} and
numerical simulations
\citep{gent1976,vanduin1992,li2000,meirink2000,sullivan2000,sullivan2008,yang2013dynamic},
there are still some very simple questions that remain only partly
answered.  One of these questions is how the form drag, exerted by the
waves on the air, depends on the properties of the air flow and the
waves.

Naturally, early attempts at numerical modelling of this flow were
devoted to the use of Reynolds averaged Navier-Stokes (RANS) models
for monochromatic waves \citep{gent1976,vanduin1992}.  The results
turned out to be sensitive to the choice of turbulence closure scheme
\citep{vanduin1992}, and suitable closures were adopted
\citep{mastenbroek1996,li2000,meirink2000}.  Increased computing
capabilities led to the advent of direct numerical simulation (DNS)
studies \citep{sullivan2000,kihara2007,yang2009,yang2010}. DNS studies
have provided valuable insight into the shape of the wave correlated
motion \citep{sullivan2000} and the structure of the turbulence
\citep{yang2009,yang2010}. Due to the strict resolution requirements
inherent to DNS, only low Reynolds numbers have been considered so
far.  \cite{meirink2000} used RANS models suitable for low Reynolds
number flows to discuss the Reynolds number effects on the drag. They
found reasonable agreement with \cite{sullivan2000} at low Reynolds
number, but further showed that there are considerable differences
between the form drag obtained at low and intermediate Reynolds
numbers. Large eddy simulation (LES) studies have primarily been aimed
at closing the gap to operational models, with special attention given
to subgrid near-surface modelling and the inclusion of a realistic
wave spectrum \citep{sullivan2008,yang2013dynamic,hara2015}. However,
\cite{yang2013dynamic} used wall resolved LES at an intermediate to
high Reynolds number as a baseline for their discussions on subgrid
modelling.

Waves that are slower than the bulk wind extract momentum from the
wind through the action of the wave-correlated pressure against the
slope of the surface. These waves are responsible for most of the
momentum transfer from the atmosphere to the sea. In this part of the
spectrum, breaking is frequent, which increases both drag and aerosol
transport \citep{donelan1998air}.  Operational wave forecasting relies
on the celebrated critical layer mechanism of
\cite{miles1957,miles1959}, where the discontinuity in wave-correlated
Reynolds stresses at the critical layer is responsible for
resonance. The existence of the critical layer, and the corresponding
jump in wave-correlated Reynolds stresses, has been demonstrated both
numerically \citep{sullivan2000} and in field observations
\citep{hristov2003}. However, \cite{belcher1993,belcher1998} argued on
the basis of a rapid distortion framework, that the Miles mechanism
could only be active for relatively fast waves and reintroduced a
non-separated version of the \cite{jeffreys1925} sheltering mechanism
for slow waves.

Waves that propagate faster than the bulk wind typically emerge as the
result of nonlinear interaction of slower waves, or as the result of
distantly storm-generated waves that enter domains of lighter
winds. The momentum flux between the air and the sea is dramatically
altered in the presence of fast moving waves
\citep{sullivan2008,kahma2016}. In severe situations, fast waves even
have the ability to generate wind \citep[see for
  instance][]{semedo2009}.  Examples of simplified models for the
air-side response in this regime include
\cite{cohen1999,kudryavtsev2004,semedo2009}.

In the present work, we perform wall resolved LES over monochromatic
waves for different wave ages and Reynolds numbers.  With wall
resolved LES the requirements for resolving the turbulence in an
adequate manner are less severe than in DNS, and as such, it may be a
valuable tool for studying the flow dynamics in the intermediate
Reynolds number range.  The layout of this paper is as follows:
Section \ref{sec:prob} contains the problem formulation and a
description of the mathematical tools used. In section \ref{sec:les}
we analyse the LES results in terms of Reynolds number and wave age
dependence. Both the mean and the fundamental mode momentum balances
are investigated to quantify the relevance of the different stresses
involved.  The LES results point towards the existence of a
quasi-laminar regime at intermediate to high wave ages, characterised
by a balance between viscous and wave induced stresses. Therefore, in
section \ref{sec:linear}, we tentatively perform a splitting of the
flow field into a laminar wave-generated response \citep{lamb1932},
and a turbulent shear flow.  The former is driven by the
non-homogeneous boundary conditions, whereas the latter is driven by a
constant volume force, turbulent stresses and the laminar solution.
We approach the problem in the opposite direction compared to for
instance \cite{miles1957}, where the wave-correlated motion is found
by linearisation about an estimated mean flow.  The aim of the present
formulation is to emphasise the role of the simple wave induced motion
for intermediate to high wave ages and explore the possibility of
linear solutions in the resulting wave correlated flow fields.  With
linear solutions, we mean linear in the sense of a flat boundary
layer; while turbulent fluctuations are active at all scales, we
investigate whether or not the wave-correlated mean flow associated
with the turbulent shear flow interacts with itself.

\section{Problem formulation}\label{sec:prob}
\begin{figure}
\centering{
\includegraphics[width=0.8\linewidth]{\figdirp 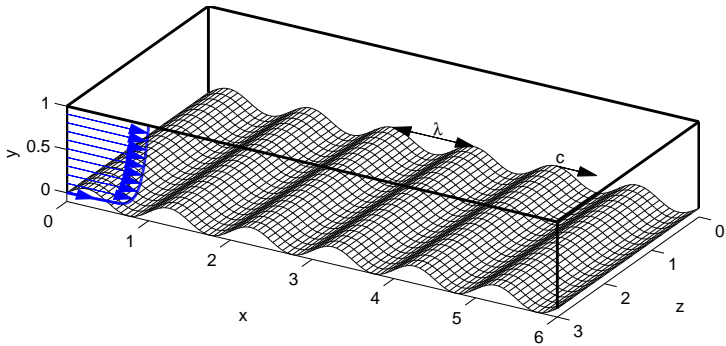}}
\caption{Illustration of computational domain.}
\label{fig:comp_box}
\end{figure}

A fully developed pressure-driven open channel flow with a moving
lower wavy surface is considered.  Figure \ref{fig:comp_box} gives an
illustration of the set up. The domain size is $(L_x,L_y,L_z) =
(6,1,3) $ wave lengths in the streamwise, vertical and crosstream
directions, respectively. The lower surface is described by a
propagating wave $y=\eta(x,t)=a\sin (k(x-ct))$ with amplitude $a$,
wave number $k=2\pi/\lambda$, and phase velocity $c$. At this
boundary, the flow is subject to kinematic boundary conditions from
the wave orbital movement imposed by the propagating wave. In the
frame of reference that moves with the propagating wave, the wave
shape and the wave orbital velocities are stationary,
\ie,
\begin{equation} (u,v,w)_{y=\eta(x)} =
  (akc\sin(kx)-c,-akc\cos(kx),0).
\end{equation}
At the top boundary,
slip conditions are prescribed, whereas in the streamwise and
crosstream directions periodicity is imposed.  The fluid motion is
described by the incompressible Navier--Stokes equations
\begin{subequations}\label{eq:navier}
\begin{align}
\deriv{u_i}{t} + u_j \deriv{u_i}{j} &= -\deriv{p}{i} + \deriv{2 \nu_t
  s_{ij}}{j} + \frac{\utauv^2}{h}
\delta_{1i},\label{eq:naviera}\\ \deriv{u_i}{i} & =
0\label{eq:navierb},
\end{align}
\end{subequations}
where $x_i=(x_1,x_2,x_3)=(x,y,z)$ denotes the streamwise, vertical and
crosstream coordinates, respectively. The velocities in the
corresponding directions are $u_i = (u_1,u_2,u_3) = (u,v,w)$.  In the
above equations, the summation convention applies to repeated indices.
The constant density is absorbed into the pressure, $p$, and the
volume force, $\utauv^2/h$. This volume force is defined as in
\cite{kihara2007}, using a virtual friction velocity $\utauv$ at the
surface. As in that work, we keep $\utauv=1$ for all Reynolds numbers
and wave ages, and the actual computed friction velocity, $\utau$, as
given later in equation \eqref{eq:utau_def}, is almost identical to
this virtually defined friction velocity for all cases
considered. Furthermore, the traceless strain-rate tensor is $s_{ij} =
1/2 (\deriv{u_i}{j}+\deriv{u_j}{i})$ and we have allowed for a varying
kinematic viscosity, $\nu_t(x_i,t)$, since the equations will be
solved using LES to account for unresolved turbulence.
    
\subsection{Averaging, RANS and Fourier transform} \label{Fourier_Section}
For analysis, the simulation results are averaged in the crosstream
direction and time, thus enabling the Reynolds decomposition
\begin{equation}
u_i(x_j,t) = \av{u_i(x_j,t)} + u'_i(x_j,t) = U_i (x,y) +
u'_i(x_j,t),\quad p(x_j,t) = P(x,y)+p'(x_j,t).
\end{equation}
For simplicity, throughout the rest of the paper we will let $(u,v,p)$
refer to the mean velocity field $(U(x,y)+c,V(x,y),P(x,y))$ and
$(u'_i,p')$ refer to the turbulent fluctuations. In the frame of
reference moving with the wave, the Reynolds Averaged Navier-Stokes
formulation of \eqref{eq:navier} is
\begin{subequations}\label{eq:nsrans}
\begin{align}
  (u-c) \deriv{u}{x} + v \deriv{u}{y} & = - \deriv{p}{x} + \nu
  \nabla^2 u - \deriv{\av{u'u'}}{x} - \deriv{\av{u'v'}}{y} +
  \utauv^2/h, \label{eq:nsransu}\\ (u-c) \deriv{v}{x} + v \deriv{v}{y}
  & = - \deriv{p}{y} + \nu \nabla^2 v - \deriv{\av{u'v'}}{x} -
  \deriv{\av{v'v'}}{y}, \label{eq:nsransv}\\ \deriv{u}{x} +
  \deriv{v}{y} = 0, \label{eq:nsransdiv}
\end{align}
\end{subequations}
where we have used the same averaging as above, and $\av{u'_i u'_j}$
are the Reynolds stresses. In the above equations, the subgrid eddy
viscosity terms have not been included, because their contribution to the  the momentum balances considered in the rest of the paper is small.

To aid the analysis, the RANS equations are Fourier transformed in the
streamwise direction. This may be achieved by first mapping the
physical curved domain into a rectangular domain by the transformation
\begin{equation}\label{eq:geom_transform} \chi = x, \qquad \xi = y - g(\chi,\xi) =
  y - \frac{ai}{2}(\exp(-ik\chi) - \exp(ik\chi))\exp(-k\xi).
\end{equation}
In this coordinate system the RANS equations \eqref{eq:nsrans} can be
approximated by
\begin{subequations}\label{eq:rans_geom}
\begin{align}
(u-c) \deriv{u}{\chi} + v \deriv{u}{\xi} & = -\deriv{p}{\chi} + \nu
  \nabla^2_\chi u - \deriv{\av{u'u'}}{\chi} - \deriv{\av{u'v'}}{\xi}
  +\utauv^2/h +\L_u + \nu\L_N u,\\ (u-c) \deriv{v}{\chi} + v
  \deriv{v}{\xi} & = -\deriv{p}{\xi} + \nu \nabla^2_\chi v -
  \deriv{\av{u'v'}}{\chi} - \deriv{\av{v'v'}}{\xi} + \L_v + \nu\L_N
  v,\\ \deriv{u}{\chi} + \deriv{v}{\xi} & = g_\chi \deriv{u}{\xi},
\end{align}
\end{subequations}
where
\begin{equation} \nabla^2_\chi=\mmderiv{}{\chi}{}+\mmderiv{}{\xi }{}, \qquad
\L_N= g_\chi^2 \mmderiv{}{\xi} - 2 g_\chi \mnderiv{}{\chi}{\xi} + (
g_\chi g_{\chi\xi} - g_{\chi\chi}) \mderiv{}{\xi}, \end{equation}
and
\begin{equation} \L_u = (u-c) g_\chi \deriv{u}{\xi} + g_\chi
  \deriv{p}{\xi} + g_\chi \deriv{\av{u'u'}}{\xi}, \qquad \L_v = (u-c)
  g_\chi \deriv{v}{\xi} + g_\chi \deriv{\av{u'v'}}{\xi}.
\end{equation}
Here, we have used the approximation $\deriv{}{y} =
(1-g_\xi)^{-1}\deriv{}{\xi} \approx \deriv{}{\xi}$. Strictly, this
requires a lower wave steepness than what is considered here, but for
analysis purposes it is more convenient to use the simplified
expression. In section \ref{sec:linsol}, where a numerical solution to
the system is sought, the full expression is used.

The variables are expanded in multiples of the fundamental mode,
\begin{equation}\label{eq:fourtrans} f(\chi,\xi) = \sum_{m=-M}^{M} \hat{f}_m(\xi)
  \exp(i k m \chi), \quad \hat{f}_{-m} = \hat{f}_m^*,
\end{equation}
where $\hat{f}$ denotes the complex Fourier amplitude function and $M$
is the number of modes.  The fundamental mode is $m=\pm 1$, the purely
real zero component, or mean, is $m=0$, and higher harmonics are $m =
\pm 2, \pm 3, \ldots$.

Nonlinear advection terms and terms involving $g_\chi$, henceforth
referred to as geometric terms due to their link to the surface shape,
both give rise to convolution terms. Note specifically that advection
terms involving $\deriv{}{x}$ transforms to
\begin{equation}\label{eq:dx_transform} \mathcal{F} (f \p_x h) \simeq
\mathcal{F}(f \p_\chi h - f g_\chi \p_\xi h) =
\mathcal{F}(f)*(\mathcal{F}(\p_\chi h)-\mathcal{F}(g_\chi \p_\xi h)).
\end{equation}
Since the function $g(\chi,\xi)$ is periodic with wave number k, the 
Fourier transform of the last term leads to a coupling with higher and 
lower harmonics,
\begin{equation}\label{eq:dygeom} \mathcal{F}(g_\chi \p_\xi h)_m = \geomfac \dy
(\hat{h}_{m-1} + \hat{h}_{m+1})=\dy_g (\hat{h}_{m-1} + \hat{h}_{m+1})=
  \dy_g [\hat{h}_m].
\end{equation}
Throughout the rest of the paper the $\xi-$derivative will be denoted
$\deriv{}{\xi} = \dy$ and the shorthand notation for the geometric
coupling contribution $\dy_g [\hat{h}_m]$ will be used.  The
continuity equation \eqref{eq:nsransdiv} for mode $m$ is then simply
\begin{equation}\label{eq:ransfourdiv}
ik m\, \hat{u}_m - \dy_g [\hat{u}_m] + \dy \hat{v}_m = 0,
\end{equation}
and the $m$-th mode of an advection term can be expressed as
\begin{multline}
  \mathcal{F}(f \p_\chi h - f g_\chi \p_\xi h)_m =\hat{f}_0
  \left(imk\hat{h}_m-\dy_g [\hat{h}_m]\right)
  +\\ \hat{f}_{-1}\left(i(m+1)k\hat{h}_{m+1}-\dy_g
  [\hat{h}_{m+1}]\right) + \hat{f}_{1}\left(i(m-1)k\hat{h}_{m-1}-\dy_g
  [\hat{h}_{m-1}]\right) + \ldots
\end{multline}
By application of the above two relations, the momentum equations
\eqref{eq:nsransu} and \eqref{eq:nsransv} can be transformed to
\begin{align}
    ik m\,(\wuu_m -c\hat{u}_m) + \dy \wuv_m &= i k m \left( i k m \nu
    \hat{u}_m -\hat{p}_m - \tuu_m \right) + \dy \left(\nu \dy
    \hat{u}_m - \tuv_m\right) + \nonumber\\
    & \utauv^2/h\, \delta_{0m}
    + \dy_g [\tuu_m + \hat{p}_m
      +\wuu_m-c\hat{u}_m] \label{eq:ransfouru} \\
    ik m\,(\wuv_m
    -c\hat{v}_m) + \dy \widehat{vv}_m &= ikm\,\left(\nu ikm \hat{v}_m
    - \tuv_m\right) + \dy \left( \nu \dy \hat{v}_m - \hat{p}_m -
    \widehat{\av{v'v'}}_m\right) +\nonumber\\
    & \dy_g \left[\tuv_m
      +\wuv_m-c\hat{v}_m\right]
    \label{eq:ransfourv}
\end{align}
These equations have been casted to conservative form in order to
simplify the identification of the various stresses in the flow.
Since the wave-correlated stresses are nonlinear terms, they couple
higher and lower harmonics. This can be seen from the expansion of the
wave correlated shear stress
\begin{equation} \label{eq:decomp}
\wuv_m = \mode{u}{0}\mode{v}{m} + \mode{u}{-1}\mode{v}{m+1} +
\mode{u}{1}\mode{v}{m-1} + \mode{u}{-2}\mode{v}{m+2 }
+\mode{u}{2}\mode{v}{m-2} + \ldots
\end{equation}
The geometrical correction, $\dy_g$, to the streamwise derivative also
introduces a coupling between higher and lower harmonics, even for
linear terms such as pressure, as seen from equation
\eqref{eq:dygeom}.  In equation \eqref{eq:ransfouru} and
\eqref{eq:ransfourv}, the geometry corrections to the Laplacian have
been omitted.  We have verified numerically that their contributions
to the analysis performed in subsequent sections are negligible.

There is a difference in the way the wave-correlated stresses and the
Reynolds stresses are treated. The wave-correlated stresses are
inherently mean flow quantities. Therefore, these stresses are
obtained as products of the individual Fourier components of the mean
flow. On the other hand, the Reynolds stresses are turbulent
quantities and their individual modes, $\widehat{\av{u'_i u'_j}}_m$,
can only be accessed from a Fourier transform of $\av{u'_i u'_j}$.

\section{Large Eddy Simulation}\label{sec:les}
\subsection{Numerical method}
The Navier--Stokes equations \eqref{eq:navier} are solved by means of
LES, using the unstructured finite-volume node-based solver VIDA from
Cascade Technologies \cite[see for
  instance][]{ham2006,ham2007towards}. LES subgrid terms are modelled
by means of a dynamic Smagorinsky procedure suitable for unstructured
grids, as described in \cite{mahesh2004}. The equations are advanced
in time using the second order BDF-2 scheme, and a fractional step
predictor-corrector procedure is employed to ensure conservation of
mass. The computational grid is generated by first considering a
rectangular domain with coordinates $(\chi,\xi,\zeta)$.  Uniform
spacing is imposed in the two horizontal directions ($\chi$ and
$\zeta$) and a grid stretching is applied in the vertical direction
($\xi$). For all cases, the average grid spacing in viscous units
along the surface is $\Delta \xi^+=\Delta \xi_{\mathrm{min}}\utau/\nu
=0.7\lp$.  A transformation to physical space is obtained by means of
\begin{equation}\label{eq:smoothness}
  x = \chi, \quad y = \eta(\chi) + \xi (1-\eta(\chi)/h), \quad z =
  \zeta,
  \end{equation}
 where $\xi \in [0,h]$, and $\eta(\chi)$ is the surface deformation.
 This results in an $\mathcal{O}(a/h)$ variation of the grid spacing
 over a wavelength close to the surface. Furthermore, since there is
 surface stress variation over a wavelength, the effective viscous
 grid spacing close to the surface varies from $\Delta y^+\approx 0.2$
 to $\Delta y^+\approx1.5$. Note that this transformation is different
 from the one used in to transform the equations for analysis.  The
 grid spacing is comparable to previous DNS studies
 \cite[]{sullivan2000,kihara2007,yang2010}.  However, we use a second
 order finite volume framework, which for the same grid spacing is not
 expected to resolve the physics as accurately as the pseudo-spectral
 methods used in previous DNS studies
(see for instance \cite{boyd2001} for differences in global versus local approximation methods).
Therefore, we have used LES to account for unresolved turbulent fluctuations.

\begin{table}
\begin{center}
\begin{tabular}{ccccccccc}
$Re_\tau$ & $\Delta x \lpi$  & $\Delta z \lpi$ & $\Delta \xi_{\mathrm{max}} \lpi $ &$N_x$& $N_z$& $N_y$ & $\cp$ \\ 
\hline
200 & 8.0 & 8.0 & 7.1 & 150 & 75 & 70 & 0,2,4,8,12,16,24,36 \\
260 & 8.7 & 7.8 & 8.1 & 200 & 100 & 85 & 4 \\
395 & 9.5 & 7.9 & 12.4 & 250 & 150 & 94 & 0,2,4,8,12,16,24,36 \\
950 & 9.5 & 9.5 & 27.4 & 600 & 300 & 130 &2,4,8,12,16,24 \\
\end{tabular}
\end{center}
\caption{Computation cases. All cases use a geometric stretching in
  the vertical direction of $r=1.03$ and the spacing from the surface
  to the first point in the domain is $\Delta \xi_{\mathrm{min}} \lpi
  = 0.7$.}\label{tab:simcases}
\end{table}

The flow over the waves depend on both the wave friction Reynolds
number, $\retau = \ut \lambda/\nu$, henceforth referred to as the
Reynolds number, and the wave age, $c/\utau$. The simulations cover a
range of wave ages for the three Reynolds numbers
$\retau=\{200,395,950\}$. In addition, a simulation at $\cp=4$ with
$\retau = 260$ was performed in order to compare with the shear-driven
flow of \cite{sullivan2000}. Table \ref{tab:simcases} shows the
details of the different simulations.

\subsection{Mean flow description} \label{sec:meanFlow}
\begin{figure}
\centering{
  \includegraphics[width=0.9\linewidth]{\figdirp 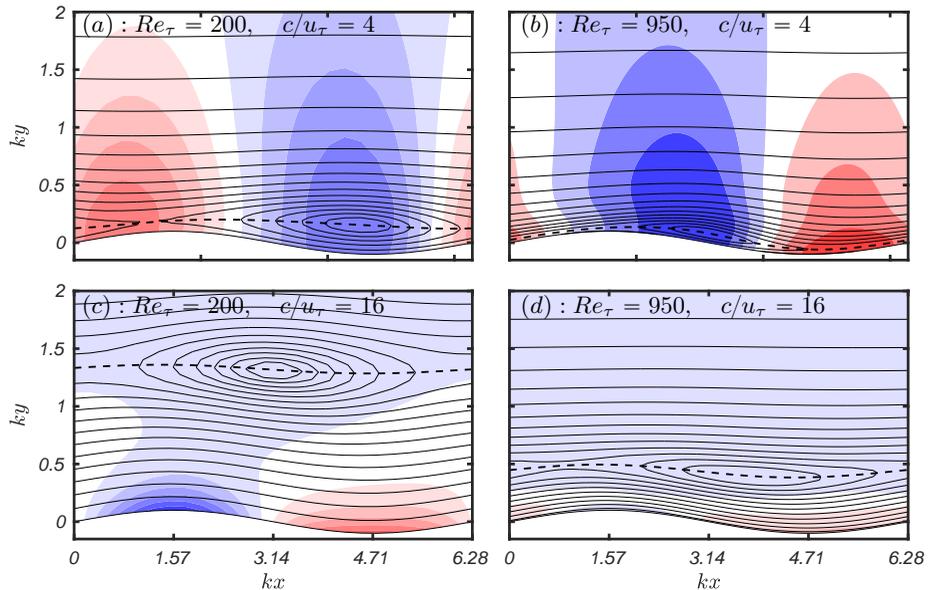}}
\setlength{\unitlength}{1mm}
\begin{picture}(0,0)(0,0)
  \put(-114,73.5){\color{black}$(a): \retau=200, \quad \cp=4$}
  \put(-56,73.5){\color{black}$(b): \retau=950, \quad \cp=4$}
    \put(-114,36.5){$(c): \retau=200, \quad \cp=16$}
  \put(-56,36.5){\color{black}$(d): \retau=950, \quad \cp=16$}
\end{picture}
\caption{Colour contours of pressure $akp/\utau^2$ and streamlines of
  the mean flow (\thinline) for (a) $\retau=200$ at $\cp=4$, (b)
  $\retau=950$ at $\cp=4$, (c) $\retau=200$ at $\cp=16$ and (d)
  $\retau=950$ at $\cp=16$.  The critical layer, \ie where $u = c$, is
  shown as (\linedasht). The colouring ranges from $akp/\utau^2=-0.6$
  (dark blue) to $akp/\utau^2=0.6$ (bright red) with an increment of
  $0.2$.  The zero contour is located between white and
  blue. }\label{fig:contours}
\end{figure}
An illustration of how the flow field varies with Reynolds number and
wave age is found in figure \ref{fig:contours}. It contains pressure
contours and mean flow streamlines for combinations of two wave ages
($\cp=4$ and $16$) and two Reynolds numbers ($\retau=200$ and $950$).
The critical layer, \ie where the streamwise velocity $u$ matches the
wave speed $c$, is shown as (\linedasht).  Above (below) this line,
the mean flow is faster (slower) than the wave.  Around the critical
layer, the so called \cite{kelvin1880} cat's-eyes appear, as seen by
the closed streamlines.  The cat's-eye plays an integral role in the
theory of \cite{miles1957}, and its role in turbulent flows has been
discussed in \cite{sullivan2000,kihara2007}.

An increase in wave age shifts the location of the cat's-eye outwards.
On the other hand, an increase in Reynolds number compresses the mean
flow profile towards the surface as the boundary layer becomes
steeper.  In viscous units, however, the critical layer position is
almost constant for low wave ages. Specifically, for the wave age
$\cp=4$, the critical layer location is $\xi \utau/\nu\approx 5$ for
all Reynolds numbers considered. For higher wave ages, the critical
layer location increases with Reynolds number, and for $\cp=16$ its
location ranges from $\xi \utau/\nu\approx 40$ for $\retau=200$ to
$\xi\utau/\nu\approx60$ for $\retau=950$.  Furthermore, the vertical
extent of the cat's-eye decreases with Reynolds number, thus
approaching the inviscid form assumed in \cite{miles1957}.

The form drag is a result of asymmetry in the pressure about the wave
crest.  Specifically, a positive (negative) form drag is present when
the low pressure is behind (in front of) the crest. This is the case
for both Reynolds numbers at the low wave age, with a more optimal
distribution (for generating form drag) and higher amplitude in the
high Reynolds number case.  At this wave age, the critical layers are
close to the surface and appear highly correlated with the pressure
minima.  This suggests that the interaction between the background
mean flow and the wave motion is dynamically important. For the
intermediate wave age cases this correlation is no longer present, and
the surface pressure variation appears to be contained in a region
well beneath the critical layer. The pressure distribution is also
more or less symmetric, resulting in low form drag. Consequently, the
variation in form drag with Reynolds number is small at intermediate
wave ages.
\begin{figure}
\centering{
  \includegraphics[width=0.8\linewidth]{\figdirp 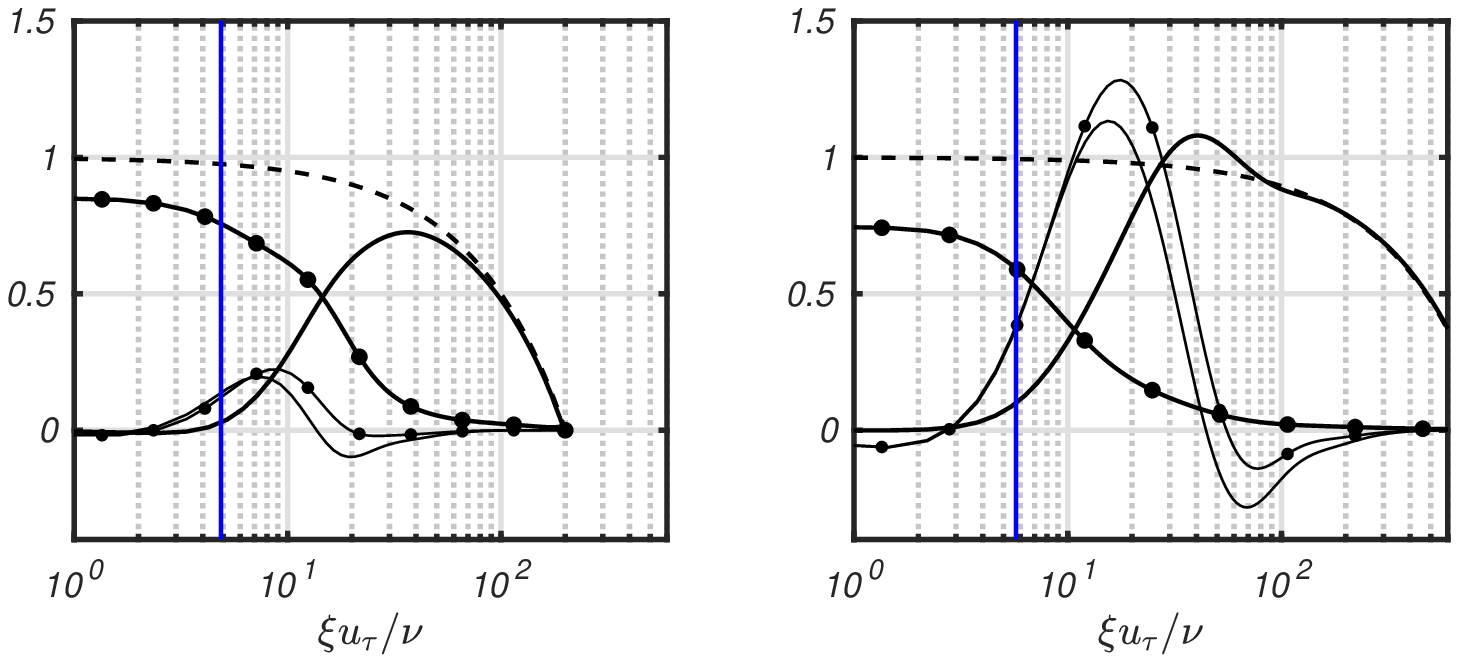}\\
    \includegraphics[width=0.8\linewidth]{\figdirp 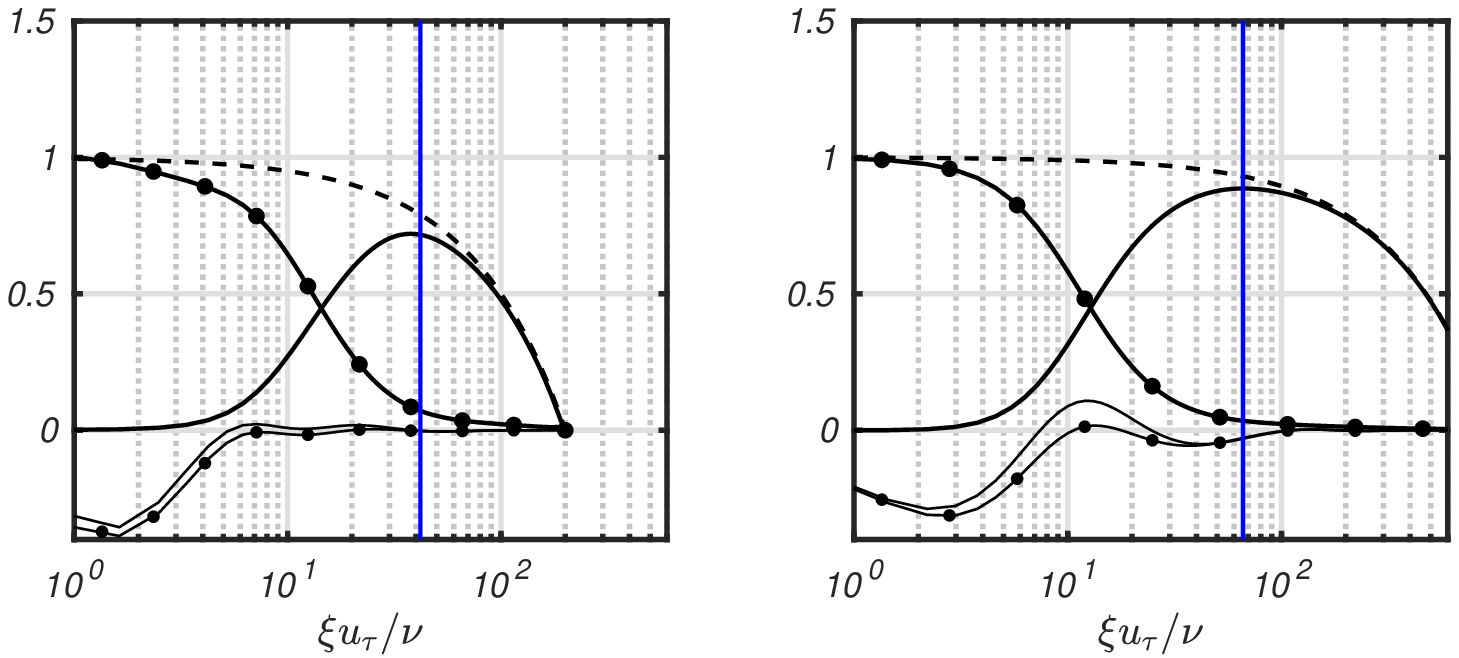}}
\setlength{\unitlength}{1mm}
\begin{picture}(0,0)(0,0)
\put(-115,94){$(a)$}
  \put(-57,94){$(b)$}
\put(-115,45){$(c)$}
  \put(-57,45){$(d)$}
\end{picture}
\caption{Vertical profiles of selected terms involved in the mean
  streamwise momentum balance for (a) $\retau=200$ at $\cp=4$, (b)
  $\retau=950$ at $\cp=4$, (c) $\retau=200$ at $\cp=16$ and (d)
  $\retau=950$ at $\cp=16$.  Reynolds shear stress $-\tuv_0/\utau^2$
  (\thickline), viscous stress $\retau^{-1} \dy \hat{u}_0 \, h/\utau$
  (\ccirclinet), channel flow balance term $1-\xi/h$ (\linedasht),
  wave induced shear stress $\wuv_0$ (\ccircline) and premultiplied
  wave induced streamwise stress $ak
  \exp(-k\xi)(\wuu_{1}^r-c\hat{u}_1^r)/\utau^2$ (\thinline). The
  location of the critical layer, where $\hat{u}_0(\xi)=c$, is shown
  as (\bthickline).}\label{fig:zero_balance}
\end{figure}

To see how the wave-correlated motion affects the mean flow balance,
we consider the streamwise momentum equation \eqref{eq:ransfouru} for
the zero mode
\begin{multline}\label{eq:zeromode}
    \dy \wuv_0 -  2\dy_g( \wuu_{1}^r -  c \hat{u}_1^r) = \dy (-\widehat{\langle u'v'\rangle}_0 + \nu\dy \hat{u}_0 ) + \utauv^2/h + 2\dy_g (\tuu_1^r + \hat{p}_1^r),
\end{multline}
where, as before, $\dy_g=\frac{1}{2} ak \exp(-k\xi) \dy$ is the
geometric correction associated with the streamwise derivative.  The
various contributions to the wave-induced stresses can easily be
obtained from the decomposition in equation \eqref{eq:decomp}.  In
standard channel flow, there is a balance between the first three
terms on the right-hand side of equation \eqref{eq:zeromode}
\citep[see for instance][]{pope2000}.  The presence of the wave
modifies this balance, resulting in additional wave-induced stresses
on the left-hand side of the equation, as well as geometry induced
turbulence and pressure terms on the right-hand side.  The first term
on the left-hand side contains the wave correlated shear stress as
defined in \cite{hussain1970,hussain1972a,hussain1972b}. However,
\cite{hussain1970} explicitly subtracted the mean before forming their
wave correlations. In the present work, the zero mode contribution is
retained in the stress.  We believe that it should be considered as
part of the wave-induced stress, since it is a result of the wave
motion and surface geometry.

To illustrate the dependence of the various stresses on Reynolds
number and wave age, we plot their respective values, for the same
cases as in figure \ref{fig:contours}, in figure
\ref{fig:zero_balance}.  Since the streamwise wave-induced stress in
equation \eqref{eq:zeromode} is not on a form that allows direct
comparison with the shear stresses, we have approximated its influence
$ak \exp(-k\xi) \dy (\wuu_1^r-c\hat{u}_1^r)$ as $\dy (ak\exp(-k\xi)
(\wuu_1^r-c\hat{u}_1^r))$. Consequently, the streamwise wave-induced
stress is presented in premultiplied form. Note that the wave-induced
streamwise and shear stresses have opposite signs in the
equation. Hence, in the figure, it is the difference of these stresses
that yields their dynamical relevance.  For all cases, the turbulent
shear stress is dominant in the outermost part of the flow, and the
momentum balance is similar to that of a channel flow. For the low
wave age cases, there is a remarkable dependence in the amplitude of
the wave-induced stresses with Reynolds number. For $\retau=200$, they
have moderate amplitudes and peak in a region where viscous stresses
are dominant. On the other hand, for $\retau=950$ the wave-induced
stresses are large in most of the flow domain. They peak well outside
the viscous sublayer, and interact with the turbulent shear stress, as
seen from the peaks in both terms at $\xi\utau/\nu\approx50$. For the
intermediate wave age case, the situation is quite different. The
wave-induced stresses have similar amplitudes for the two Reynolds
numbers, and their imbalance primarily occurs in the viscous sublayer.
There thus appears to be less interaction between turbulent and
wave-induced stresses as the wave age increases.

\subsection{Form drag}
\begin{figure}
\centering{ \includegraphics[width=0.9\linewidth]{\figdirp 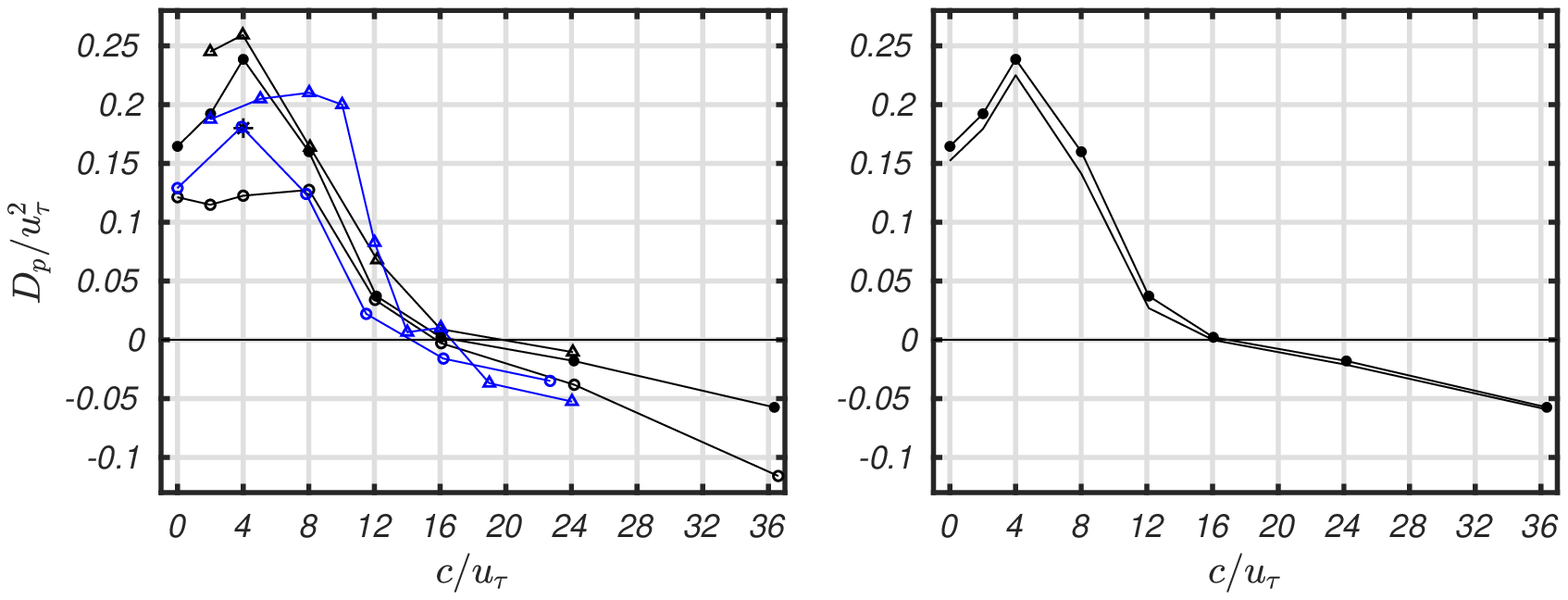}}
\setlength{\unitlength}{1mm}
\begin{picture}(0,0)(0,0)
\put(-120,45){$(a)$}
\put(-60,45){$(b)$}
\end{picture}
\caption{ Form drag as function of wave age $\cp$.  (a) Current
  simulations for Reynolds number $\retau=200$ (\ocircline),
  $\retau=260$ ($\convolution$), $\retau=395$ (\ccircline),
  $\retau=950$ (\triline) . Couette flow DNS of \cite{sullivan2000} at
  $\retau=260$ ({\color{blue}\ocircline}) and $\retau=1000$ RANS
  results ({\color{blue}\triline}) of \cite{meirink2000}. The
  $ak=0.01$ results of \cite{meirink2000} has been transformed from a
  growth rate factor $\beta$ to $D_p$ using $ak=0.1$.  (b) Comparison
  of form drag at $\retau=395$ using the surface pressure (\ccircline)
  and the integral of the wave induced shear stress (\thinline) as
  defined in equation \eqref{eq:presbalance}.}
\label{fig:form_drag}
\end{figure}
Figure \ref{fig:form_drag}(a) shows a comparison of the form drag,
$D_p / \ut^2$, obtained from the LES at different Reynolds numbers as
a function of wave age. In addition, the low Reynolds number DNS
results of \cite{sullivan2000} and the intermediate Reynolds number
RANS results of \cite{meirink2000} are also plotted.  The form drag,
$D_p$, is defined as the pressure contribution to the surface stress
\begin{equation}\label{eq:utau_def}
u_\tau^2 = \frac{1}{\lambda } \int_0^\lambda \left(\nu
(\deriv{u}{y}+\deriv{v}{x} - 2 \deriv{u}{x}\deriv{\eta}{x}) + p
\deriv{\eta}{x}\right) \dint x = D_v + D_p.
\end{equation}

For wave ages above $\cp\approx 10$ there is a small upward shift in
form drag with increasing Reynolds number for the current
simulations. The same behaviour is seen in \cite{meirink2000}.  This
implies that the wave age at which the form drag switches from
positive (wave growth regime) to negative (wave decay regime) shifts
towards higher wave ages.  All results point towards a low Reynolds
number sensitivity in this regime.

Below $\cp\approx 10$ both the present results and other published
work show considerable variation in form drag as a function of
Reynolds number, which is consistent with the large Reynolds number
sensitivity of the wave-induced stresses discussed in section
\ref{sec:meanFlow}.  Before we describe the differences, we first note
that the Couette flow results of \cite{sullivan2000} and
\cite{yang2010} are in good agreement, despite the large differences
in friction Reynolds number based on the channel half height,
$\retau^h = \utau h/\nu$. On the other hand, their wave friction
Reynolds numbers are almost identical.  This suggests that the wave
length, as opposed to the half channel height, is the appropriate
outer length scale in this flow.  This observation is further
supported by comparing our $\retau^h = \retau= 200$ results with the
results of \cite{kihara2007}, where $\retau^h=150$ and
$\retau=200$. Indeed, our results are in excellent agreement with
\cite{kihara2007}.

At $\retau=200$ the maximum form drag is found at $\cp=8$, but in
contrast to the higher Reynolds numbers, there is relatively small
wave age sensitivity below $\cp\approx10$. Although the Reynolds
number in the shear-driven flow of \cite{sullivan2000} is only $30\%$
larger, there is a distinct difference between the two at $\cp=4$. In
fact, the results of \cite{sullivan2000} resemble our $\retau=395$
results, although with a slightly lower form drag. At first glance,
this seems to support the well known observation that shear-driven
flows inherently contain more high Reynolds number dynamics than a
pressure-driven flow at the same Reynolds number.  To check this, we
performed a simulation at $\retau=260$ for this wave age. As seen from
the figure, our form drag is almost identical to the one in
\cite{sullivan2000}.  Therefore, it appears that the large variation
across data sets at low wave ages is not primarily a flow
configuration issue. Instead, the results point towards a high
Reynolds number sensitivity at low wave ages.

Both at $\retau=395$ and $950$ there is a distinct peak in form drag
at $\cp=4$. The $\retau=1000$ ({\color{blue}\triline}) results of
\cite{meirink2000} has a peak at $\cp=8$, but their variation with
wave age in this regime is small. They also found that maximum form
drag was obtained at $\retau=800$, and linked this to an optimal
cooperation of viscous and turbulent stresses. For $\cp=5$ they also
report a higher form drag at $\retau=260$ (figure 9 in that paper)
than at $\retau=1000$. This is clearly contrasted by our $\cp=4$
results, where $\retau=950$ has a substantially higher form drag than
$\retau=260$.  Our results do not contradict the observed $\retau=800$
maximum in \cite{meirink2000}, but the agreement between our
$\retau=950$ results and their $\retau=1000$ results is rather poor.
Note also that their $\retau=260$ results are only partly in agreement
with \cite{sullivan2000}.  Although \cite{meirink2000}, with support
from \cite{gent1976}, claim that the growth rate is independent of
wave steepness below $ak<0.1$, we are not convinced that this is the
case.

 \begin{figure}
  \centering{ \includegraphics[width=0.9\linewidth]{\figdirp
      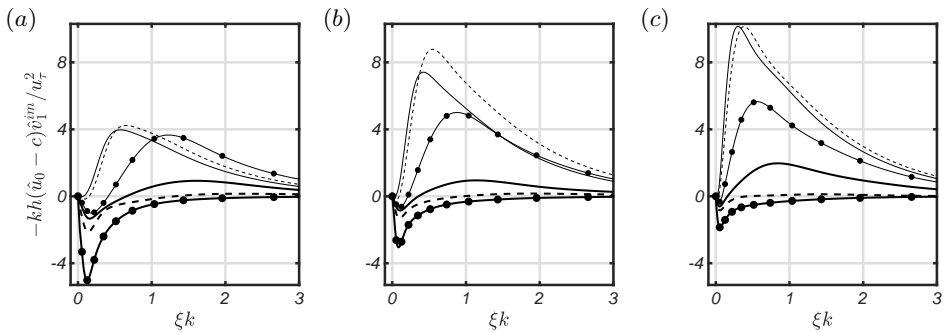}} \setlength{\unitlength}{1mm}
  \begin{picture}(0,0)(2,3)
    \put(-124,43){$(a)$}
    \put(-82,43){$(b)$}
    \put(-40,43){$(c)$}
  \end{picture}
  \caption{Vertical profiles of the fundamental mode shear stress term
    $-kh(\hat{u}_0-c)\vo$ for Reynolds numbers (a) $\retau=200$, (b)
    $\retau=395$ and (c) $\retau=950$. The different wave ages are
    $\cp=2$ (\thinline), $\cp=4$ (\linedash), $\cp=8$ (\ccircline)
    $\cp=12$ (\thickline), $\cp=16$ (\linedasht) and $\cp=24$
    (\ccirclinet). Note that only approximately half of the vertical
    domain is shown.}\label{fig:wave_stress_fund}
\end{figure}

 From equation \eqref{eq:utau_def} we see that the form drag is a
 result of the action of the pressure against the slope of the
 surface. Since the slope is given as
 \begin{equation}
   \eta_x = 0.5 a
   k(\exp(ikx)+\exp(-ikx)),
 \end{equation}
 we end up with the simple relation,
 \begin{equation}
 D_p = \lambda^{-1}\int_0^\lambda \eta_x p \dint
 x = \frac{1}{2} a k (\hat{p}_1 + \hat{p}_{-1}) = ak \hat{p}_1^{r}.
 \end{equation}
From this we observe that the form drag is determined by the
out-of-phase pressure at the surface.  We follow
\cite{mastenbroek1996,meirink2000}, where the out-of-phase surface
pressure component is found by integrating the vertical momentum
equation \eqref{eq:ransfourv} from the surface to the freestream. They
neglected nonlinear and geometric terms, which could be justified by
their low wave steepness.  Even for the present steepness of $ak=0.1
$, we find that the surface pressure to a large extent may be computed
from the wave-induced shear stress
component \begin{equation}\label{eq:presbalance} \hat{p}_1^r \approx
  \left\{\int_{0}^{1} ik (\hat{u}_0 -c) \hat{v}_1 \dint
  \xi\right\}_{re}=\int_{0}^{1} -k(\hat{u}_0 -c) \hat{v}_1^{im} \dint
  \xi,
\end{equation}
with the remaining contribution mainly stemming from the turbulent
stress component $\av{v'v'}$. A comparison of the pressure drag
computed using equation \eqref{eq:presbalance} and the LES results can
be found in figure \ref{fig:form_drag}(b).
 
 The vertical dependence of the wave-induced shear stress,
 $-k(\hat{u}_0 -c) \hat{v}_1^{im}$, for different Reynolds numbers and
 wave ages, is seen in figure \ref{fig:wave_stress_fund}.  For wave
 ages below $\cp=12$, all Reynolds numbers have predominantly positive
 wave-induced shear-stress profiles, resulting in positive form drag.
 For $\cp=16$, there is a negative region close to the surface which
 is approximately balanced by a positive outer region, resulting in
 low form drag. For the highest wave age, $\cp=24$, all Reynolds
 numbers have strictly negative profiles.  As the wave age increases,
 the support of the wave-induced stresses are increasingly confined
 close to the surface. The Reynolds number dependence is also
 simplified in the sense that the shape becomes similar. The amplitude
 decreases with increasing Reynolds number. On the other hand, for low
 wave ages the wave-induced stress has support throughout the boundary
 layer and the dependence of peak locations and amplitudes are
 non-monotonic in the parameters.
 
\subsection{Reynolds number dependence at a low wave age}\label{sec:lowwaveage}
\begin{figure}
  \centering{
        \includegraphics[width=0.75\linewidth]{\figdirp 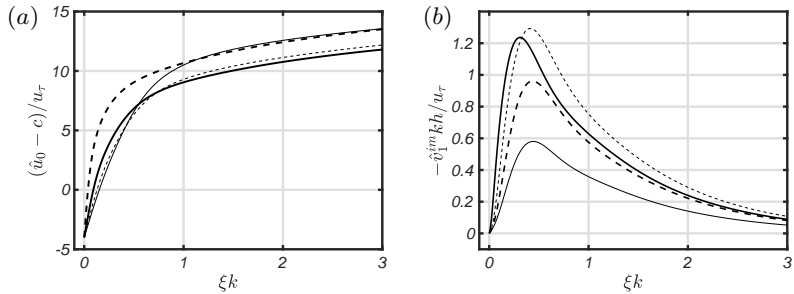}}
  \setlength{\unitlength}{1mm}
  \begin{picture}(0,0)(2,4.5)
    \put(-104,40){$(a)$}
    \put(-49,40){$(b)$}
  \end{picture}
  \caption{Vertical profiles of terms contributing to the fundamental
    mode wave-induced shear stress at $\cp=4$ for Reynolds numbers
    $\retau=200$ (\thinline), $\retau=260$ (\linedash), $\retau=395$
    (\thickline), $\retau=950$ (\linedasht). (a) Streamwise mean
    velocity $(\hat{u}_0-c)/\utau$. (b) Out-of-phase vertical velocity
    $-kh\,\hat{v}_1^{im}/\utau$. Note that only approximately half the
    vertical domain is shown.}
  \label{fig:pressure_contrib}
\end{figure}

To gain insight into the high Reynolds number sensitivity at low wave
ages, we consider the wave age $\cp=4$, where simulations have been
performed at all four Reynolds numbers.  Figure
\ref{fig:pressure_contrib}(a) and (b) show the mean flow
$(\hat{u}-c)/u_\tau$ and the out-of-phase vertical velocity
$-kh\hat{v}_1^{im}/u_\tau$, respectively.  Close to the surface, the
mean flow profiles display the expected behaviour of increasing
gradients with increasing Reynolds number. In the outer part of the
flow, however, there is no evidence of a direct link between the form
drag and the mean flow magnitude. In other words, a lower freestream
velocity does not follow from an increased form drag, as would be the
case for standard channel flow.  This may seem counter intuitive, but
as seen from the presence of wave-induced stresses in the zero mode
streamwise momentum balance \eqref{eq:zeromode}, there is no a priori
reason for one to follow from the other. Naturally, given the
increasing form drag with Reynolds number, the out-of-phase vertical
velocity also exhibits a non-monotonic behaviour with increasing
Reynolds number.  For all Reynolds numbers, the out-of-phase vertical
velocity has support up to $\xi k \approx 3$.  It therefore has the
ability to interact with the mean flow in a large part of the flow
domain.

Overall, the Reynolds number dependence can not be attributed to the
individual behaviour of the mean flow or the out-of-phase vertical
velocity, since both of these vary in a non-monotonic manner. In order
to assess which physical processes that are involved in determining
the wave-induced stresses, we consider the streamwise momentum balance
for the fundamental mode,
\begin{multline}\label{eq:fund_mom_balance}
 \underbrace{\dy (c\hat{v}_1 + \wuv_1) + ik\,\wuu_1 - \dy_g [\wuu_1]}_{\textrm{Wave induced}}  =
\underbrace{\nu \dy^2  \hat{u}_1}_{\textrm{Viscous}} - \underbrace{(\dy \tuv_1 + ik\tuu_1 - \dy_g [\tuu_1])}_{\textrm{Turbulent}} - \\\underbrace{(ik\hat{p}_1 + \dy_g [\hat{p}_1])}_{\textrm{Pressure}}.
\end{multline}
Here, we have used the approximation $(\dy^2-k^2) \approx \dy^2$. Specifically, we 
consider the imaginary part of this equation, since it contains $\hat{u}_0 \hat{v}_1^{im}$.
\begin{figure}
  \centering{
        \includegraphics[width=0.9\linewidth]{\figdirp 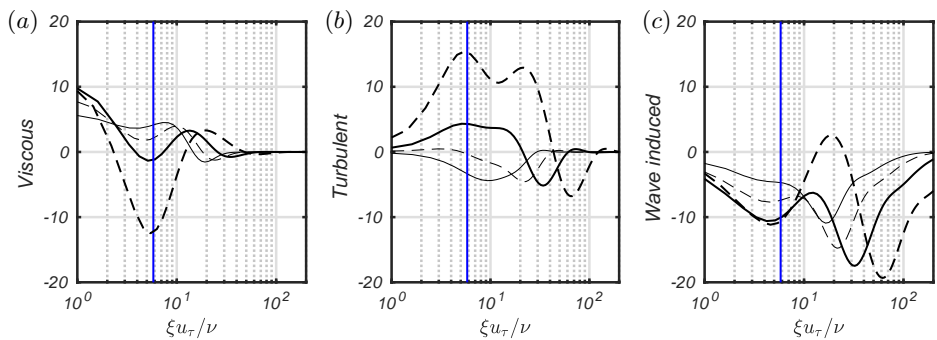}}
  \setlength{\unitlength}{1mm}
  \begin{picture}(0,0)(1,2)
    \put(-124,43){$(a)$}
    \put(-82,43){$(b)$}
    \put(-40,43){$(c)$}
  \end{picture}
   \caption{Vertical profiles of stress gradients in the out-of-phase
     streamwise momentum equation \eqref{eq:fund_mom_balance} for
     $\cp=4$ and $\retau=200$ (\thinline), $\retau=260$ (\linedash),
     $\retau=395$ (\thickline), $\retau=950$ (\linedasht). The
     location of the critical layer for $\retau=950$ is shown as
     (\bthinline).
} \label{fig:fund_mom_balance}
\end{figure}

Figure \ref{fig:fund_mom_balance} shows the imaginary part of (a) the
viscous, (b) the turbulent and (c) the wave-induced stress gradients
for all four Reynolds numbers. Notice that all stress terms contribute
significantly to the momentum balance even in the interior of the
domain. The contribution of the pressure is almost constant in most of
the domain, and it is thus not shown in the figure.

For all Reynolds numbers, the viscous forces are active from the
surface well beyond the critical layer.  While the $\retau=200$ case
exhibits an almost constant behaviour below the critical layer, an
increasing variation in the viscous force is seen with increasing
Reynolds number.  For $\retau=950$ there is a negative peak of similar
amplitude as the surface value. This indicates that for the
fundamental mode, the importance of viscosity around the critical
layer increases with Reynolds number.

The turbulent stress gradients display a large variation with Reynolds
number. It is interesting that the turbulent and viscous stress
gradients have opposite signs and similar amplitudes around the
critical layer for all cases.  Given the small variation in the
wave-induced stresses in this part of the domain, the results hint at
a balance between viscous and turbulent forces close to the surface.
In the figure, the turbulent stress gradient is the sum of the
streamwise and shear-stress contributions. However, at the critical
layer, the shear stress is much larger than the streamwise stress for
$\retau=950$.  With decreasing Reynolds number, the streamwise stress
becomes more important. This type of behaviour can be explained by
comparing the streamlines in figure \ref{fig:contours}(a) and (b). For
$\retau=200$ the vertical extent of the cat's eyes region is
large. Therefore, a fluid particle located in its centre will
experience a smaller shear anisotropy than in the $\retau=950$ case.

In the outer part of the flow, the amplitudes of the turbulent and
wave-induced stress gradients increase with Reynolds number in a
correlated manner. Notice in particular the extremal points for
$\retau=950$ around $\xi \utau/\nu\approx 20$ and $\xi \utau/\nu=60$.
This behaviour supports the rapid-distortion theory of
\cite{belcher1993}.

\subsection{Reynolds number dependence at an intermediate wave age}
We next consider the case $\cp=16$. At this wave age, all simulations
have close to zero form drag. Previously it was conjectured that the
flow exhibits a simpler behaviour in this regime.  From figure
\ref{fig:wave_stress_fund} we also observed that the out-of-phase
wave-induced stress term $(\hat{u}-c)\hat{v}_1^{im}$ behaved similarly
for all Reynolds numbers.

\begin{figure}
  \centering{
    \includegraphics[width=0.9\linewidth]{\figdirp 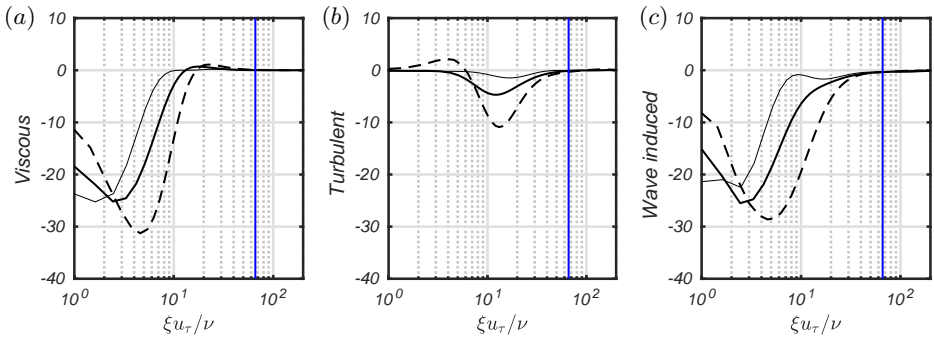}}
  \setlength{\unitlength}{1mm}
  \begin{picture}(0,0)(1,2)
    \put(-124,43){$(a)$}
    \put(-82,43){$(b)$}
    \put(-40,43){$(c)$}
  \end{picture}
   \caption{Vertical profiles of stress gradients in the out-of-phase
     streamwise momentum equation \eqref{eq:fund_mom_balance} for
     $\cp=16$ and Reynolds numbers $\retau=200$ (\thinline),
     $\retau=395$ (\thickline), $\retau=950$ (\linedasht). The
     location of the critical layer for $\retau=950$ is shown as
     (\bthinline).} \label{fig:fund_mom_balance_16}
\end{figure}

To investigate why this is the case, we consider the stresses in the
out-of-phase streamwise momentum equation as a function of Reynolds
number. The resulting stress gradients are found in figure
\ref{fig:fund_mom_balance_16}. At this wave age, all the dynamics in
the fundamental mode occur well within the critical layer, and the
main contributing terms are the viscous and wave-induced stresses. The
turbulent stress gradient becomes more pronounced with increasing
Reynolds number, and for $\retau=950$ it has a significant
contribution above $\xi\utau/\nu=10$. The dominance of viscous and
wave-induced stresses is even more pronounced at $\cp=24$ and $36$. It
thus seems that a quasi-laminar regime is entered in the intermediate
to high wave age regime.

Note that the wave age can be written as a ratio of Reynolds numbers
\begin{equation}\label{eq:cpdef}
\frac{c}{\ut} = \frac{c}{k\nu} \frac{\nu}{\ut \lambda} (2\pi) =
\frac{Re_w}{\retau} (2 \pi),
\end{equation}
where $Re_w$ is the wave Reynolds number \citep{lamb1932}. A large
wave age implies that the wave Reynolds number is large compared to
the friction Reynolds number, or that the smallest scales in the flow
are governed by the imposed wave. It thus seems that the simple wave
age dependence at higher wave ages is the result of a separation of
scales. Turbulence mainly has its dynamical relevance in maintaining
the mean flow, while the flow response to the propagating wave is
maintained by a near-surface balance between viscous and wave-induced
forces. These observations, coupled with the seemingly simple
dependence of $D_p$ on $Re_\tau$ at intermediate to high wave ages,
suggest that splitting the flow fields into a wave response and a
remainder may be fruitful.

\section{A split system approach}\label{sec:linear}
Consider a splitting of the flow fields into a ``shear'' contribution
$(u_i^h,p^h)$ and a wave contribution $(u_i^p,p^p)$. The latter,
hereafter referred to as the particular solution, satisfies the
non-homogeneous Dirichlet boundary conditions imposed by the
travelling wave, and is the solution to Navier-Stokes equations in the
absence of a background shear flow.  The former, hereafter referred to
as the homogeneous solution, is subject to homogeneous boundary
conditions, and is driven by the particular solution and the constant
body force, as well as turbulent stresses. The system is hence built
in a bottom-up approach (from the known boundary conditions), instead
of the common top-down approach
\citep{miles1957,belcher1993,kudryavtsev2001}, where linear solutions
about a prescribed estimate of the mean flow are sought.

To illustrate the splitting of the flow fields, we consider the
decomposed nonlinear advection terms
\begin{equation}
u_j\deriv{u_i}{j} =
\underbrace{u_j^p\deriv{u_i^p}{j}}_{\textrm{particular}} +
\underbrace{u^p\deriv{u_i^h}{j} + u_j^h\deriv{u_i^p}{j}
  +u_j^h\deriv{u_i^h}{j}}_{\textrm{homogeneous}}.
\end{equation} 
The first advection term is solved separately as the laminar flow
response to the non-homogeneous boundary conditions, whereas the three
latter terms are parts of the homogeneous system that is forced by the
particular solution.  It is expected that the velocities of the
particular solution scale as $ack$, and that the homogeneous solution
has a mixed scaling, involving both $ack$ and $\ut$. We may not a
priori state the importance of the different terms, but linearity in
the homogeneous system equations would result from the terms
$u_j^h\deriv{u_i^h}{j}$ being small compared to the others.

\subsection{The particular solution}
For the particular solution we use a Helmholtz decomposition
\begin{equation}
\up = \upot + \us,
\end{equation}
where $\upot=\deriv{\phi}{i}$ is irrotational and $\us$ is divergence
free.  The velocity potential can be obtained by solving the Laplace
equation $\deriv{\phi}{ii} = 0$ subject to $n_i \deriv{\phi}{i} = n_i
u_i$ at the surface and $\deriv{\phi}{y} =0$ in the freestream. Here,
the surface velocity $u_i$ is given by linear Airy wave theory. The
equation for the steady streamfunction velocity is
\begin{equation}\label{parteq}
-c \deriv{\us}{x} + \upotj \deriv{\us}{j} + \usj \deriv{\upot}{j}
+\underbrace{\usj \deriv{\us}{j}}_{\textrm{nonlinear}}
= -\deriv{p_{\psi}}{i} + \nu
\deriv{\us}{jj}, \quad \deriv{\us}{i} = 0,
\end{equation}
where $\us = (u^\psi,v^\psi)$ is subject to the boundary conditions
$\us = u_i - u_i^\phi$ at the lower boundary $y=\eta$, and $\us = 0$
in the freestream.  When the nonlinear terms are skipped, the above
equations represent a linear system forced by a periodic base flow
with wave number $k$. The Helmholtz decomposition enables interaction
between the velocity potential and the streamfunction velocity through
the advection terms. An interesting observation is that the
introduction of the geometry transformation \eqref{eq:geom_transform}
leads to the following cancellation
\begin{equation}
-c \deriv{\us}{x}+ v^\phi \deriv{\us}{y} = - c\deriv{\us}{\chi} + c
 g_\chi (1+g_\xi)^{-1} \dy{\us} -v^\phi (1+g_\xi)^{-1} \dy{\us} = -
 c\deriv{\us}{\chi},
\end{equation}
where we have used $c g_\chi = v^\phi$. This implies that the single
mode behaviour to leading order is well described by a linear system
with $\us =\epsilon_{ij3}
\deriv{\psi}{j}=(\dy{\psi},-\deriv{\psi}{\chi})$,
\begin{equation}\label{stream}
-c \nabla_\chi^2 \psi_\chi  = \nu \nabla_{\chi}^4 \psi, 
\end{equation}
subject to $\dy{\psi} = -2 ack \sin(k\chi)$ and $\deriv{\psi}{\chi}=0$
at the lower surface $\xi=0$, and $\dy{\psi}=\deriv{\psi}{\chi}=0$ at
the top boundary $\xi=1$. This system is the one-phase version of the
system in \cite{lamb1932,harrison1908}, where we match only the
kinematic properties of the lower phase. Due to the viscous scaling of
the streamfunction velocities, the standard linear wave theory
approximation of evaluating at $y=0$ is prone to increasing errors
with increasing wave Reynolds number.  Therefore, the boundary
conditions must be evaluated at the interface ($\xi=0$).

The main motivation for resorting to a coupled formulation is that we
are concerned with the multi-mode behaviour of the particular
solution.  Specifically, we are dealing with the transfer of momentum
from the fundamental mode to lower and higher harmonics, which is
enabled by the action of the velocity potential on the viscous
solution. The particular solution will in turn act as a multi-modal
base flow for the homogeneous solution, and the effect of small mean
or higher harmonic contributions can not be neglected a priori.

\subsection{The homogeneous solution}
Once the particular solution is obtained, one may solve for the
homogeneous solution, which has the following RANS formulation
\begin{equation}\label{homeq}
-c \deriv{\uh}{x} + \upj \deriv{\uh}{j} + \uhj \deriv{\up}{j} +
\underbrace{\uhj \deriv{\uh}{j}}_{\textrm{nonlinear}} =
-\deriv{p_{h}}{i} + \nu \deriv{\uh}{jj} + \utauv^2/h \delta_{1i} -
\deriv{\av{u_i' u_j'}}{j},\quad\deriv{\uh}{i} = 0.
\end{equation}
The velocity field is subject to a no-slip condition at the surface,
and a slip condition in the freestream. Once nonlinear terms ($\uhj
\deriv{\uh}{j}$) are skipped, these equations represent a linear
system forced by a multi-modal base flow.  The body force term,
$\utauv^2/h$, is chosen in accordance with the LES set up, and the
turbulent Reynolds stresses are taken from the LES simulations.

The choice of using turbulent stresses from the LES renders this
framework of little practical use as a RANS model, and indeed this is
not the aim of the present work. Instead, we aim at using the split
system approach to analyse the LES results and better understand the
underlying dynamics of the flow. It should be pointed out that we also
could have obtained the homogeneous solution by subtracting the
particular solution from the LES results directly, but this approach
is not as convenient for exploring the dependencies of the equation
system on various assumptions.

There has been some debate regarding how to properly model turbulent
stresses in flow over waves. \cite{belcher1993} argued, on the basis
of a rapid-distortion framework, that eddy viscosity type closures are
suitable only in the near surface region, since turbulent eddies far
from the surface are transported too quickly to be strained by the
local shear.  This has been confirmed numerically by
\cite{mastenbroek1996} who compared different RANS closures to
experimental results. In the present work, we have tested the simple
eddy-viscosity closure of \cite{reynolds1967}, based on the
\cite{cess} framework, which has proven successful in studying linear
energy amplification in turbulent channel flow \cite[]{pujals2009}.
Indeed, we found that no simple relation between the mean shear and
the turbulent stresses is present.

\subsection{Solution procedure}\label{sec:linsol}
The equation systems \eqref{parteq} and \eqref{homeq} have a similar
form that can be written
\begin{equation}\label{linsys}
\underbrace{ \ccolvec{ \mathcal{L} - U_x & -U_y & -\deriv{}{x}} {-V_x
    & \mathcal{L} - V_y & -\deriv{}{y}}{\deriv{}{x} & \deriv{}{y} &
    0}}_{A} \underbrace{\ccolvec{u}{v}{p}}_q +
\underbrace{\textrm{diag}(-u\deriv{}{x} - v\deriv{}{y},-u\deriv{}{x} -
  v\deriv{}{y},0)}_{N(q)} \underbrace{\ccolvec{u}{v}{p}}_q =
\underbrace{\ccolvec{f_u}{f_v}{0}}_f,
\end{equation}
where $\mathcal{L} = \nu \nabla^2 + (c-U)\deriv{}{x}-V\deriv{}{y}$,
with $\nabla^2 = \deriv{}{xx}+\deriv{}{yy}$ being the Laplacian
operator. The above system represents the equations for the particular
solution when the base flow is $(U,V) = (u^\phi,v^\phi)$ and the
forcing is zero ($f=0$). It represents the homogeneous solution when
$(U,V) = (u^p,v^p)$ and the forcing terms are
\begin{equation}\begin{split}
f_u &= -\utauv^2/h + \deriv{\av{u'u'}}{x}+\deriv{\av{u'v'}}{y} + f_b,\\ 
f_v &=  \deriv{\av{u'v'}}{x}+\deriv{\av{v'v'}}{y}. 
\end{split}
\end{equation}
Here, the additional forcing term $f_b$ is added to correct for
incompatibility of the divergence free criterion used in the LES and
the spectral collocation method, which leads to a momentum imbalance
of the LES results on the spectral collocation grid \cite[see][for a
  discussion on the conservation properties of the LES]{ham2004}. This
term is generally small, but the streamwise mean velocity in the outer
part of the domain is highly sensitive to any pointwise momentum
imbalance.

In the linear case, it is possible to solve system \eqref{linsys}
explicitly by introducing a coordinate transformation to map the
physical domain to a rectangular domain and then Fourier transforming
along the streamwise direction, as was done in previous sections.
However, both the periodic base flow and the geometrical terms lead to
coupling of the different harmonics, which makes this a tedious and
time consuming effort. Instead, we have used a more attractive
approach based on two-dimensional Fourier-Chebyshev collocation
\cite[]{weideman2000,easystab}. The physical domain $0 \le x \le
2\pi/k$ and $\eta(x) \le y \le 1$ is mapped to a rectangular
computational domain $0 \le \chi \le 2\pi/k$ and $0\le \xi \le 1$
using the transformation \eqref{eq:smoothness}, and the corresponding
derivative operators are redefined numerically to account for this
mapping.  Since the base flow is periodic, the resolution in the
streamwise direction can be kept rather low. For the particular
solution, we found $(N_x,N_y)=(8,100)$ to be sufficient at all
Reynolds numbers. The homogeneous solution has an additional forcing
from turbulent stresses, which results in stricter resolution
requirements. We found $(N_x,N_y)=(12,120)$ to be sufficient for the
cases considered.

In the nonlinear configuration, the system is solved using simple
Newton iterations. The linear solution is used as an initial
guess. For all but the highest Reynolds number, we found that
convergence could be reached in less than ten iterations.  At the
highest Reynolds number, convergence is sensitive to both initial
guess and the choice of iterative solver. We found that convergence
could be ensured by successively adding Fourier components of the
turbulent forcing.

\subsection{Discussion on the particular solution}\label{sec:part}
\begin{figure}
\centering{
  \includegraphics[width=0.8\linewidth]{\figdirp 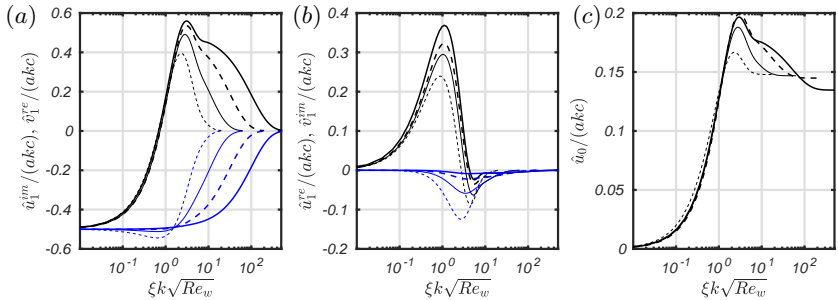}}
\setlength{\unitlength}{1mm}
\begin{picture}(0,0)(126,0)
  \put(14,36){$(a)$}
  \put(52,36){$(b)$}
  \put(89,36){$(c)$}
\end{picture}
\caption{ Streamwise (\thinline) and vertical (\bthinline) profiles of
  the particular solution as a function of distance from surface in
  viscous units.  (a) In-phase components, (b) out-of-phase components
  and (c) zero mode of the streamwise velocity.  Four wave Reynolds
  numbers, $\rew = c/k\nu$, are considered. $\rew = 10$ (\thinline\,
  and \bthinline), $\rew=10^2$ (\linedash\, and \blinedash), $\rew =
  10^3$ (\thickline\, and \bthickline), and $\rew=10^4$ (\linedasht\,
  and \blinedasht).}\label{fig:part}
\end{figure}
Figure \ref{fig:part} shows vertical profiles of the fundamental mode
velocities of the particular solution for the four wave Reynolds
numbers $Re_w = \{10,100,1000,10000\}$. To put these wave Reynolds
number in context, one may use relation \eqref{eq:cpdef}. For
instance, at $\cp=2$ the corresponding wave Reynolds numbers of
$\retau=200$ and $950$ are $Re_w\approx60$ and $300$, respectively.
The in-phase parts of the streamwise and vertical velocities are shown
in figure \ref{fig:part}(a), and the out-of-phase velocities are shown
in figure \ref{fig:part}(b).  The vertical velocity approaches the
velocity potential with increasing wave Reynolds number, which means
that the out-of-phase part tends to zero, and the in-phase part
approaches the velocity potential $-akc/2 \exp(-k\xi)$. In the outer
part of the domain, the streamwise velocity is well described by the
velocity potential, but the streamfunction velocity is needed close to
the surface to match the boundary conditions.

As previously stated, the fundamental mode of the particular solution
may be approximated from the analytical solution of equation
\eqref{stream}. For high wave Reynolds numbers ($Re_w > 100$) we may
write the solution as
\begin{equation} \psi =\{ A \left(\exp(- k \xi) - \exp(- n k
\xi)\right) \exp(ik\chi)\}_{\mathrm{re}},A = \sqrt{\frac{2}{\rew}} a c
(1-i), \quad n = \sqrt{\frac{\rew}{2}} (1-i).
\end{equation}
The velocity components are then approximately
\begin{equation}\label{simplesol} \hat{u}_1 = \dy \psi
\approx -akc i \exp(-n k \xi)\exp(ik\chi)\,\quad \textrm{and} \quad
\hat{v}_1 = - ik \psi.
\end{equation}
Here, the streamwise velocity amplitude is independent of Reynolds
number, while the vertical velocity amplitude decays with Reynolds
number as $\sqrt{2/\rew}$. We also observe that vertical profiles
collapse when scaled with $k\sqrt{\rew}$.  The same behaviour is seen
for the full particular solution, with minor corrections to the
amplitude of the streamwise velocity.  The out-of-phase streamwise
velocity reflects the adjustment from the imposed boundary condition
to the velocity potential in the outer flow.  The phase shift occurs
in the region $k\xi\sqrt{Re_w} = [0,10]$. The only clear difference
for the fundamental mode behaviour between the current formulation and
the uncoupled analytical approach given in \eqref{stream}, which is
similar the the one in \cite{lamb1932,harrison1908}, is that the peak
amplitude is slightly higher for the coupled system at high Reynolds
numbers.

We find that the coupling between the velocity potential and the
streamfunction velocity first and foremost gives rise to a positive
valued zero mode with a peak amplitude of approximately half of the
fundamental mode, as seen in figure \ref{fig:part}(c).  The amplitude
of the second harmonic is approximately $ak$ times that of the
fundamental mode for all wave Reynolds numbers considered. This is
indicative of linear behaviour. The present case, with a wave
steepness of $ak=0.1$, is transitional; for less steep waves
($ak<0.05$), coupling terms are negligible, and for steeper waves
($ak>0.15$) both coupling terms and nonlinear interactions of the
streamfunction velocity with itself become important.

\subsection{Discussion on the total solution}

\begin{figure}
\centering{
  \includegraphics[width = 0.9\linewidth]{\figdirp 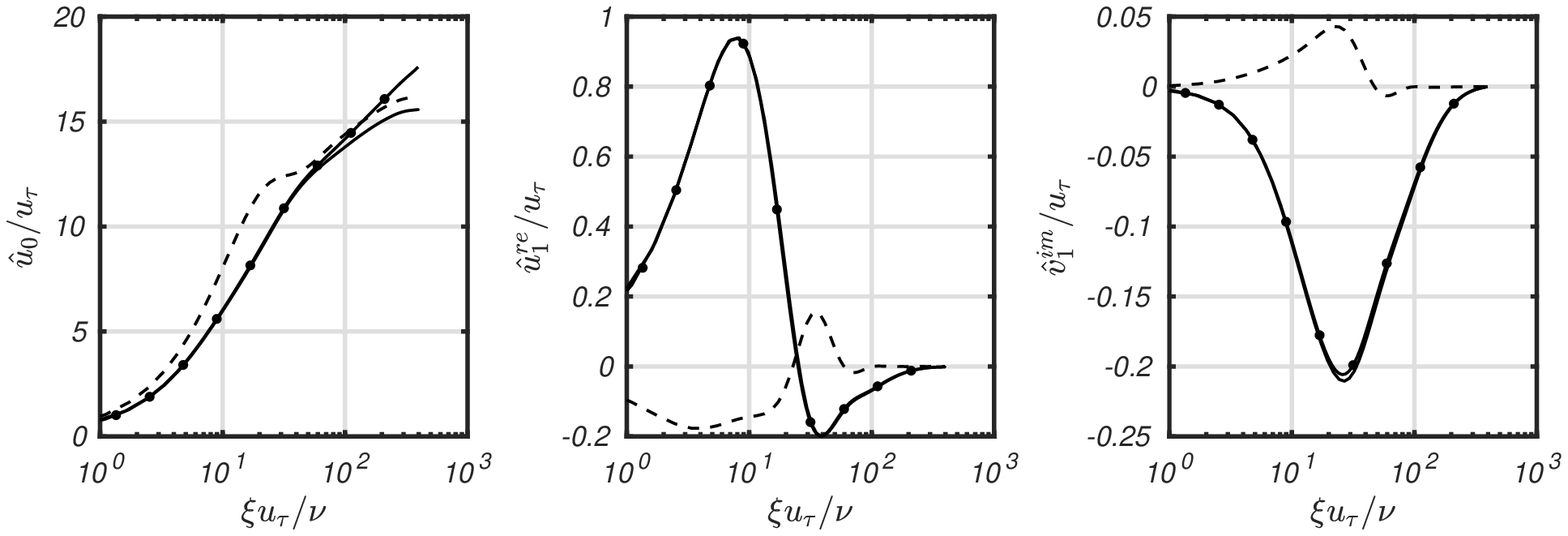}\\
  \vspace{0.3cm}
  \includegraphics[width = 0.9\linewidth]{\figdirp 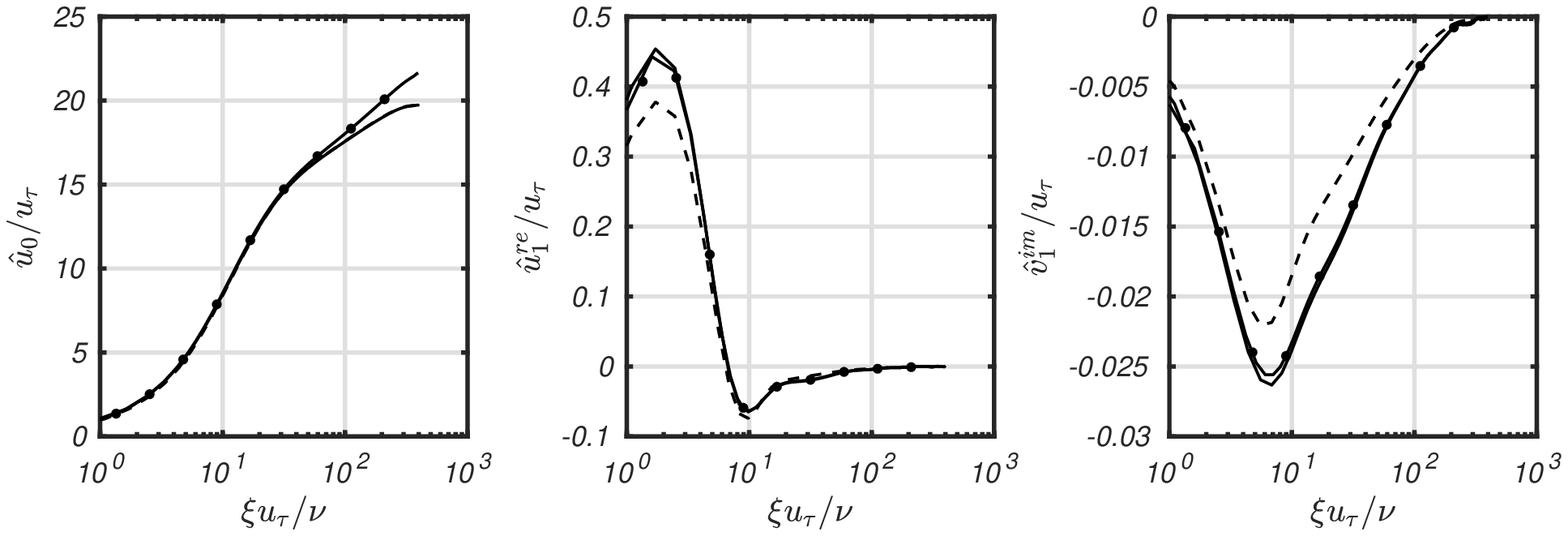}}
\setlength{\unitlength}{1mm}
\begin{picture}(100,0)(4,-2)
\put(-8,85){$(a)$}
\put(33,85){$(b)$}
\put(74,85){$(c)$}
\put(-8,40){$(d)$}
\put(33,40){$(e)$}
\put(74,40){$(f)$}
\end{picture}
\caption{Fourier-components of the total solution $u_i = \up + \uh$ of
  the split system for the linear (\linedasht) and nonlinear
  (\thickline) solution at $\retau=395$.  LES results are shown as
  (\ccirclinet). (a)-(c) shows the wave age $\cp=4$, where large
  differences are seen between the linear and nonlinear
  solution. (d)-(f) shows the wave age $\cp=24$, where the linear
  solution describes the overall features of the flow
  well.}\label{total_sol_comp}
\end{figure}

Figure \ref{total_sol_comp} shows a comparison of LES results with
both linear and nonlinear split-system solutions for $\retau=395$ at
the two wave ages $\cp=4$ and $24$. The split-system solution is the
sum of the homogeneous and particular solution. For the low wave age
case, seen in the top frames of the figure, neither the streamwise
mean velocity nor the out-of-phase fundamental mode velocities are
captured by the linear system.  On the other hand, the nonlinear
solution is in good agreement with the LES results. In the outermost
part of the domain, the streamwise mean velocity deviates
significantly from the LES. This discrepancy, which is present for all
cases considered, is most likely due to slight differences in
implementation of the top boundary condition. If we explicitly enforce
the freestream velocity from the LES as the top boundary condition,
the profiles are in close agreement. Despite the deviation of the
streamwise mean profiles in the outer part of the domain, the
agreement in form drag between the two is excellent for all wave ages
and Reynolds numbers. The reason for this, is that the deviation
occurs in a region where the out-of-phase vertical velocity has
negligible amplitude.
 
For the high wave age case, the streamwise mean velocities of the
linear and nonlinear solutions are almost identical. For the
out-of-phase velocity components, the linear solution captures the
main features of the flow, but the amplitudes are somewhat
underpredicted. For the even higher wave age of $\cp=36$, the linear
and nonlinear solutions are in excellent agreement. In the present
formulation, the linearity requirement is very strict, since it
implies that the nonlinear term $v_h \dy u_h$, which includes the mean
shear, is negligible in the momentum balance. The wave age at which
the linear regime is entered increases with Reynolds number.

\begin{figure}
\centering{
\includegraphics[width=0.42\linewidth]{\figdirp 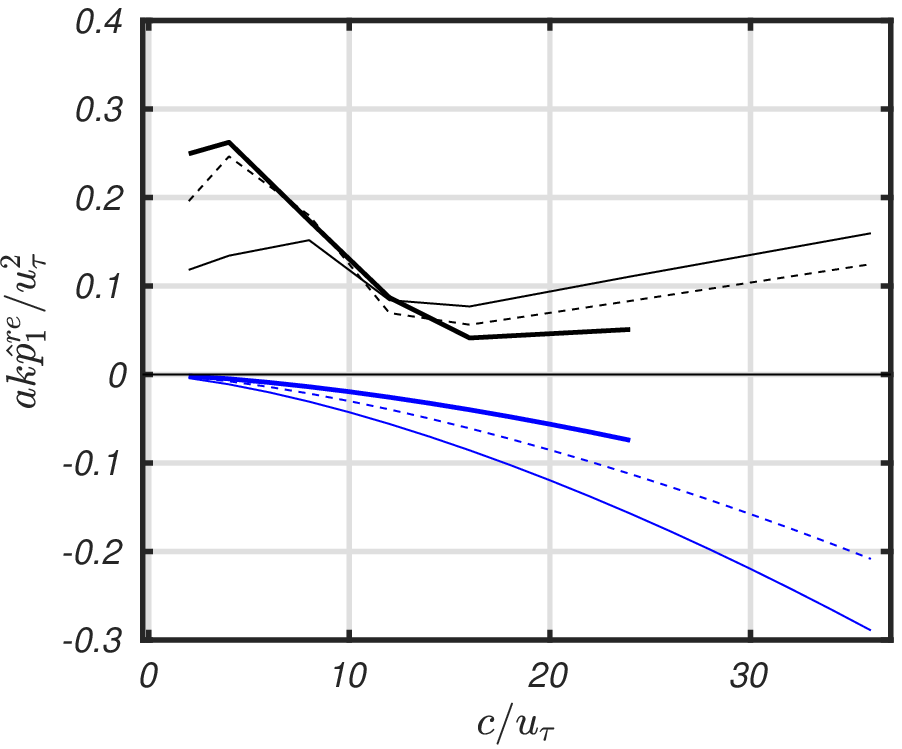}\qquad
\includegraphics[width=0.42\linewidth]{\figdirp 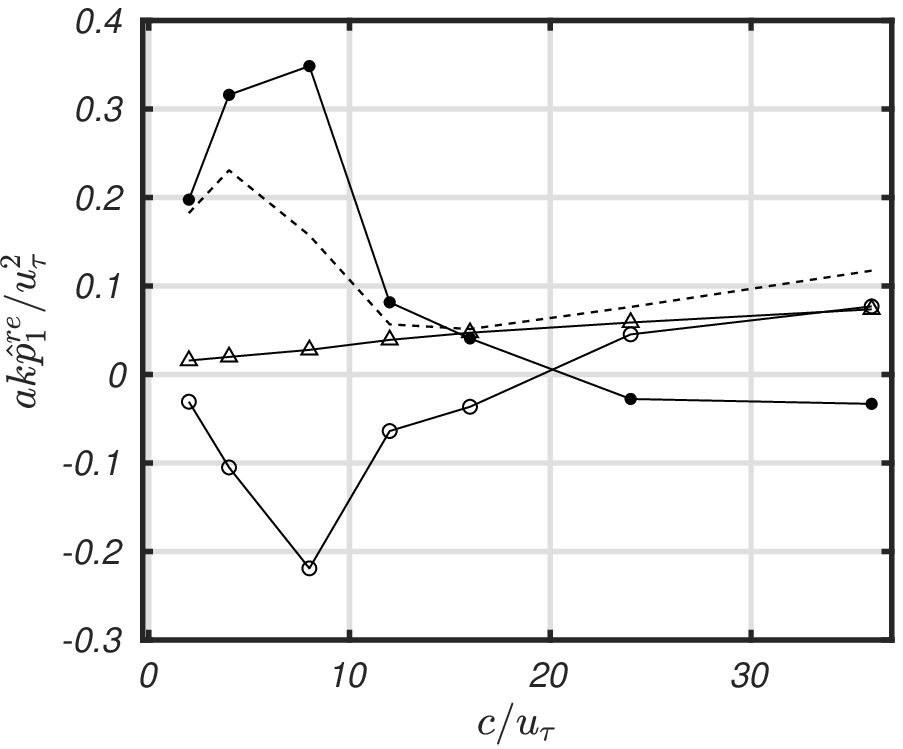}}
\setlength{\unitlength}{1mm}
\begin{picture}(0,0)(0,0)
  \put(-124,45){$(a)$}
  \put(-60,45){$(b)$}
\end{picture}
\caption{(a) Surface pressure reconstructed from the homogeneous
  solution (\thickline) and the particular solution (\bthickline) for
  the three Reynolds numbers $\retau=200$ (\thinline\, and
  \bthinline), $\retau=395$ (\linedash\, and \blinedash) and
  $\retau=950$(\thickline\, and \bthickline) at different wave
  ages. (b) Integral of terms in equation \eqref{eq:stress_decomp} at
  $\retau=395$. The contributions are $\hat{u}_0^h\hat{v}_1^p$
  (\triline), $-c \hat{v}_1^h$ (\ocircline) and
  $\hat{u}_0^h\hat{v}_1^h$ (\ccircline).  The sum of the three
  contributions is (\linedash).  }\label{fig:form_drag_system}
\end{figure}

To explore the functional dependence of the form drag with Reynolds
number and wave age, we consider the particular and homogeneous
surface pressure separately in figure
\ref{fig:form_drag_system}(a). The surface pressure of the particular
solution is strictly negative, and for low wave ages its contribution
is almost negligible compared to its homogeneous counterpart. With
increasing wave age this situation changes, and the particular
solution eventually outweighs the positive contribution of the
homogeneous solution, resulting in a net negative form drag. The
behaviour of the homogeneous pressure is non-monotonic, with a minimum
around $\cp=16$.  Beyond this, we observe an approximate linear growth
with wave age.

\begin{figure}
\centering{\includegraphics[width=0.9\linewidth]{\figdirp 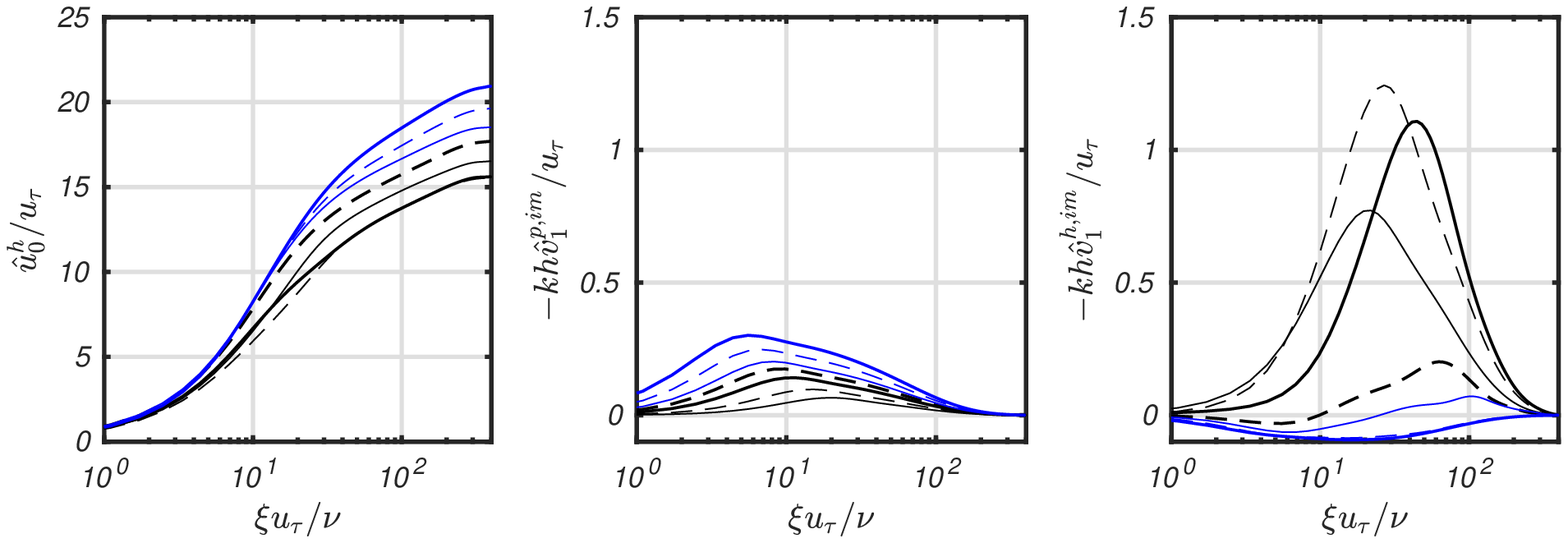}}
 \setlength{\unitlength}{1mm}
  \begin{picture}(0,0)(110,0)
   \put(-15,39){$(a)$}
   \put(27,39){$(b)$}
   \put(68,39){$(c)$}
   \put(50,9){\vector(-2,3){6}}
  \end{picture}
  \caption{Vertical profiles of velocity components involved in the
    wave induced stress at $\retau=395$ computed by means of the
    nonlinear split system.  The different wave ages are: $\cp=2$
    (\thinline), $\cp=4$ (\linedash), $\cp=8$ (\thickline), $\cp=12$
    (\linedasht), $\cp=16$ (\bthinline), $\cp=24$ (\blinedash) and
    $\cp=36$ (\bthickline).  (a) The mean streamwise velocity of the
    homogeneous solution, and the out-of-phase vertical velocity for
    (b) the particular solution and (c) homogeneous solution. The
    arrow in (b) indicates increasing wave
    age.}\label{fig:outphase_system_395}
\end{figure}

\begin{figure}
  \centering{ \includegraphics[width=0.42\linewidth]{\figdirp
      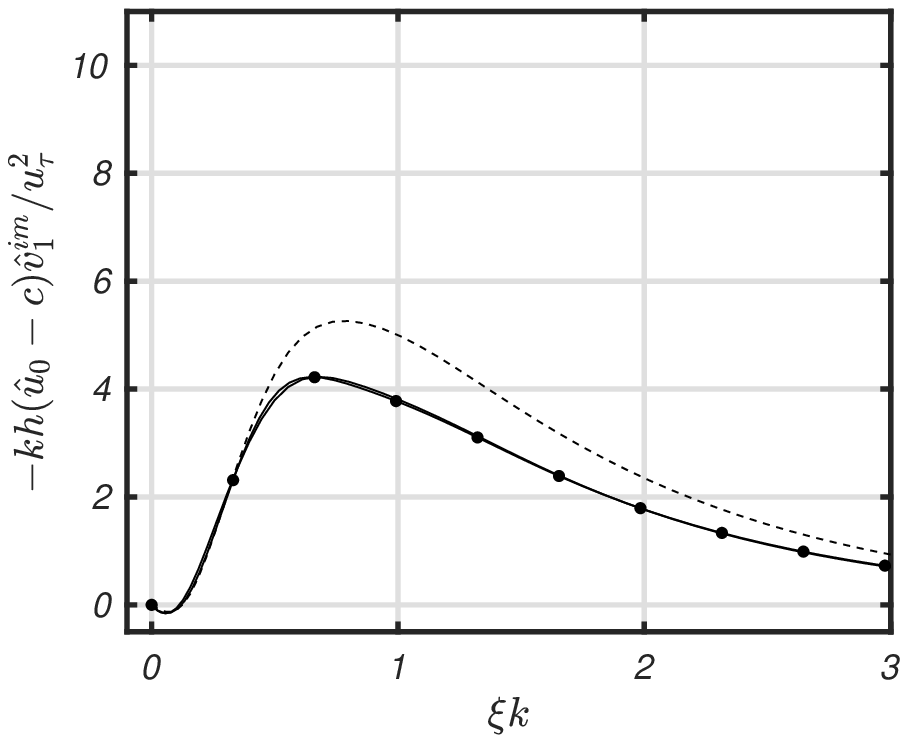}\quad \includegraphics[width=0.42\linewidth]{\figdirp
      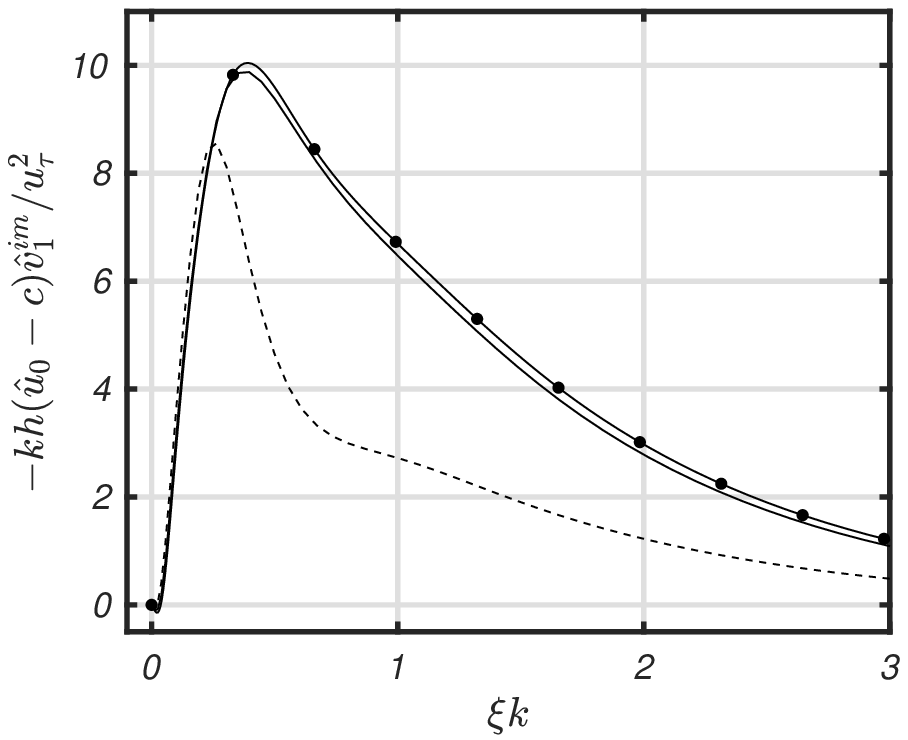}} \setlength{\unitlength}{1mm}
   \begin{picture}(0,0)(120,0)
   \put(0,44.5){$(a)$}
   \put(60,44.5){$(b)$}
   \end{picture}
  \caption{Out-of-phase shear stress from LES (\ccircline) and the
    split system approach using only the zero mode turbulent stresses
    (\linedash) and also using fundamental mode turbulent stresses
    (\thinline) for $\cp=4$. (a) $\retau=200$ and (b)
    $\retau=950$. Notice that for $\retau=200$ the LES (\ccircline)
    and the split-system solution (\thinline) is
    indistinguishable. }\label{fig:turbvar_test}
\end{figure}
As previously stated, the main contributor to the surface pressure is
the wave-induced shear stress term (see equation
\eqref{eq:presbalance}), which decomposes to
 \begin{multline} \label{eq:stress_decomp}
 - (\hat{u}_0-c)\vo kh/\utau^2 = \underbrace{(c-\hat{u}_0^p) \vpo
   kh/\utau^2}_\textrm{Particular}\\ +
 \underbrace{(c-\hat{u}_0^p-\hat{u}_0^h )\vho kh/\utau^2 -
   \hat{u}_0^h\vpo kh/\utau^2}_\textrm{Homogeneous}.
 \end{multline}
 The first term on the right hand side is the contribution to the
 particular solution pressure. Since the mean flow of the particular
 solution is at least $ak/2$ smaller than the phase velocity (see
 section \ref{sec:part}), $\hat{u}_0^p\vpo$ is small and can be
 neglected. With this simplification, a scaling for the particular
 solution pressure can be obtained using the analytical solution
 \eqref{simplesol}. The analytical solution gives the following
 estimate for the magnitude of the out-of-phase vertical velocity,
 \begin{equation}
 \vpo/\utau \sim - ak\cp Re_w^{-1/2} = - ak (2 \pi\,\cp )^{1/2}
 \retau^{-1/2}.
 \end{equation}
 This scaling is consistent with the profiles in figure
 \ref{fig:outphase_system_395}(b).  It implies that the particular
 solution pressure scales as
 \begin{equation}
 \ppo/\utau^2 \sim k h c\, \vpo/\utau^2 = - ak (kh)(2 \pi)^{1/2}
 (\cp)^{3/2} \retau^{-1/2}.
 \end{equation}
 The particular solution pressure decreases in magnitude with
 increasing friction Reynolds number and grows rapidly with increasing
 wave age. This functional form is clearly seen from the blue lines in
 figure \ref{fig:form_drag_system}(a).

 The rest of the terms on the right hand side of
 \eqref{eq:stress_decomp} contribute to the homogeneous solution
 pressure. The term containing the zero mode of the particular
 solution is small and can be neglected, whereas the three remaining
 terms are shown for $\retau=395$ in figure
 \ref{fig:form_drag_system}(b). Of the velocity components involved in
 these expressions, only the out-of-phase vertical velocity of the
 homogeneous solution displays a complicated wave age dependence (see
 fig.  \ref{fig:outphase_system_395}(c)). For low wave ages, $\vho$ is
 negative and large, with peak amplitude encountered at
 $\cp=4$. Consequently, the surface pressure is dominated by the term
 $(c-\hat{u}_0^h)\vho$. At approximately $\cp=12$ there is a switch in
 behaviour of the out-of-phase vertical velocity, and a change of sign
 occurs close to the surface. As the wave age increases, the profiles
 become strictly positive, until finally, the change with wave age is
 slow (as seen by the overlapping lines of $\cp=24$ and $\cp=36$).
 This dictates the high wave age behaviour seen in figure
 \ref{fig:form_drag_system}(b).  Firstly, the integral of
 $c\hat{v}_1^{h,\mathrm{im}}$ ends up being positive and scales
 linearly with wave age.  Secondly, the integral of
 $-\hat{u}_0^h\hat{v}_1^{h,\mathrm{im}}$ reaches a constant state,
 since $\hat{u}_0^h$ (see fig. \ref{fig:outphase_system_395}a) is
 unaltered in the region where $\hat{v}_1^{h,\mathrm{im}}$ is
 non-negligible. Lastly, the contribution from, $\hat{u}_0^h \vpo$ is
 positive and scales approximately as $(\cp)^{1/2}$. The end result is
 an almost linear dependence of the homogeneous surface pressure for
 high wave ages.  Based on the above scaling, we can estimate the
 pressure drag in the quasi-laminar regime given its inception value.
 The existence of the quasi-laminar regime is good news, as it reduces
 the parameter space of interest for future investigations to a
 smaller region of the wave age Reynolds number plane.  As previously
 noted, the onset of this quasi-laminar regime is Reynolds number
 dependent. Exploring this Reynolds number dependence is a topic for
 future work.
 
The sensitivity of the form drag to variation in Reynolds number at
low wave ages is more challenging to evaluate using the present
decomposition. From the results in figure \ref{total_sol_comp}, we
know that the full nonlinear system is required to describe the flow
dynamics. The analysis in section \ref{sec:lowwaveage} established
that the interaction of the wave-induced stress with turbulence
increases with Reynolds number. To quantify the importance of the wave
correlated turbulent stresses, we solve the nonlinear split-system
using only the zero mode of the Reynolds stresses as a forcing.  The
resulting wave-induced stresses, for $\retau=200$ and $950$ at the
wave age $\cp=4$ are found in figure \ref{fig:turbvar_test}. For
$\retau=200$, using only the zero mode yields a fairly good estimation
of the wave induced stress and the form drag is slightly
overpredicted. On the other hand, for $\retau=950$ the stress is
severely underpredicted, which results in a fifty percent
underestimation of the form drag. The fundamental mode turbulence is
thus an essential part of the dynamics responsible for the form drag.
 
Another interesting observation is that at the lowest wave age,
$\cp=2$, and $\retau=200$, using only the forcing of the zero
component yields almost identical results to the full system. In
addition, the coupling with the particular solution can be excluded
altogether, which implies that the kinematics of the underlying wave
is not important. The wave kinematics and turbulence is therefore
fully decoupled. However, as the Reynolds number is increased, the
exclusion of coupling with the particular solution leads to an
increasing inability to describe the flow. The main difference in the
particular solution with increasing Reynolds number is that the shear
becomes stronger (due to the scaling $\sqrt{Re_w}$). This means that
at low wave ages, the effect of both the laminar forcing from the wave
kinematics and turbulence increase with Reynolds number. Since the
laminar forcing also contributes to generating turbulence in multiple
harmonics, isolating the effect of one from the other is a challenge.

\section{Conclusions}
In this paper we have presented results from wall resolved large eddy
simulations of turbulent flow over a simple propagating wave at a
moderate wave steepness of $ak=0.1$. We have varied both Reynolds
number and wave age over a range previously not considered in such
detail.

A key quantity of interest is the dependence of form drag on the
governing parameters. In the intermediate to high wave age regime,
previous studies have found a somewhat simplified dependence of form
drag on both Reynolds number and wave age, but to the authors'
knowledge, conclusions have been limited to an observed collapse when
scaling the problem using an outer velocity scale instead of the
friction velocity. In the low wave age regime, the form drag depends
in a complex manner on Reynolds number, wave age and wave steepness.

We analysed the results by Fourier-decomposing the computed flow
variables, and evaluate the contributions to the momentum balance in
both the mean and fundamental mode. Using this decomposition, the form
drag can be found by integrating the vertical momentum equation from
the surface to the free stream. The dominant contribution to this
integral was found to be the wave-induced shear stress which is the
product of the streamwise mean velocity and the out-of-phase vertical
velocity. We observed that for low wave ages, this wave-induced stress
has a significant contribution in a large part of the domain. On the
other hand, at high wave ages the stress is increasingly confined to a
region close to the surface.

By examining the results in terms of Reynolds number dependence at a
low wave age, we found that the interaction between the wave induced
and turbulent stresses plays an increasingly important role in the
streamwise momentum balance of the fundamental mode as the Reynolds
number increased.  At an intermediate wave age, we found that the
fundamental mode balance was primarily dominated by viscous and wave
induced stresses. This indicates that a quasi-laminar regime, where
wave kinematics become increasingly important, is entered as the wave
age increases. Therefore, we introduce a novel split system approach,
where we first construct a laminar flow response to the wave
kinematics. This in turn acts as a forcing on a shear flow subjected
to homogeneous boundary conditions. To account for the effects of
turbulence, we force the system using Reynolds stresses from the large
eddy simulation.  By splitting the flow in this manner, we were able
to formulate an analytic functional dependence for the form drag
associated with the laminar response. The results also show that the
form drag of the shear flow simplifies at high wave ages, and an
approximate wave age dependence of the full solution could be
constructed.  The reason for this simplicity is that, at high wave
ages, the forcing from the laminar wave-induced flow overwhelms the
geometry induced interaction of the shear flow with itself.

The high sensitivity of the form drag to variation in Reynolds number
at low wave ages is more challenging to evaluate using the split
system approach. This is primarily due to the increased importance of
nonlinearity in the shear flow. These nonlinearities are inherently
coupled with higher harmonics in the turbulent stresses. Nevertheless,
the split system approach can be utilized to quantify the importance
of different harmonics in the turbulent stresses by explicitly
choosing which modes to include in the split system forcing. In
support of \cite{belcher1993}, we demonstrated that the fundamental
mode of the Reynolds stresses becomes increasingly important for
accurate prediction of the form drag as the Reynolds number
increases. However, the importance of nonlinearity in the low wave age
regime suggest that any linear low wave age model for the form drag
can only work for a very low wave steepness and/or Reynolds
number. The dependence of form drag on wave steepness and Reynolds
number in the low wave age regime is an interesting topic for future
work.

\bibliographystyle{jfm}
\bibliography{aakervik_vartdal2019}
\end{document}